\documentclass[twocolumn,showpacs,preprintnumbers,superscriptaddress,
               amsmath,amssymb,prd]{revtex4}

\usepackage[dvips]{graphicx}
\usepackage{dcolumn}
\usepackage{bm}

\newcommand{\al}{\alpha}
\newcommand{\be}{\beta}
\newcommand{\ga}{\gamma}

\newcommand{\De}{\Delta}

\newcommand{\eps}{\epsilon}

\newcommand{\Om}{\Omega}

\newcommand{\txt}{\textstyle}
\newcommand{\dsp}{\displaystyle}

\newcommand{\eqn}[1]{(\ref{#1})}  
\newcommand{\e}{{\rm e}}  
\newcommand{\beq}{\begin{equation}}
\newcommand{\eeq}{\end{equation}}
\newcommand{\ba}{\begin{array}}
\newcommand{\ea}{\end{array}}
\newcommand{\bea}{\begin{eqnarray}}
\newcommand{\eea}{\end{eqnarray}}
\newcommand{\bal}{\begin{align}}  
\newcommand{\eal}{\end{align}}
\newcommand{\bi}{\begin{itemize}}  
\newcommand{\ei}{\end{itemize}}
\newcommand{\ben}{\begin{enumerate}}  
\newcommand{\een}{\end{enumerate}}

\newcommand{\half} {{\txt \frac{1}{2}}}
\newcommand{\third}{{\txt \frac{1}{3}}}

\newcommand{\twothirds}{{\txt \frac{2}{3}}}

\newcommand\hide[1]{}

\newcommand{\Tr}{\mbox{Tr}}

\newcommand{\psibar}{{\bar\psi}}

\newcommand{\feyn}[1]{
  \setbox0=\hbox{\ensuremath{#1}}
  \hbox to\wd0{\hbox to0pt{\hbox to\wd0{\hss/\hss}\hss}\box0}}

\newcommand{\MeV}{\;{\rm MeV}}

\newcommand{\diag}{{\rm diag}}
\newcommand{\Qtilde}{{\tilde Q}}
\newcommand{\mue}{\mu_{e}}
\newcommand{\ms}{M_{s}}
\newcommand{\mssq}{M_{s}^2/\mu}

\begin{document}

\preprint{MIT-CTP-3525}

\title{Heating (Gapless) Color-Flavor Locked Quark Matter}

\author{Kenji Fukushima}
\affiliation{Center for Theoretical Physics, Massachusetts 
             Institute of Technology, Cambridge, MA 02139, USA}
\affiliation{Department of Physics, University of Tokyo,
             7-3-1 Hongo, Tokyo 113-0033, Japan}

\author{Chris Kouvaris}
\affiliation{Center for Theoretical Physics, Massachusetts 
             Institute of Technology, Cambridge, MA 02139, USA}
\author{Krishna Rajagopal}
\affiliation{Center for Theoretical Physics, Massachusetts 
             Institute of Technology, Cambridge, MA 02139, USA}

\date{August 25, 2004}

\begin{abstract}
We explore the phase diagram of neutral quark matter at high baryon
density as a function of the temperature $T$ and the
strange quark mass $M_s$.  At $T=0$, there is a sharp distinction
between the insulating color-flavor locked (CFL) phase, which occurs where
$M_s^2/\mu < 2 \De$, and the metallic gapless CFL phase, which
occurs at larger $M_s^2/\mu$.  Here, $\mu$ is the chemical potential
for quark number and $\De$ is the gap in the CFL phase.  We find
this distinction blurred at $T\neq 0$, as the CFL phase 
undergoes an insulator-to-metal crossover when it is heated.
We present an analytic treatment of this crossover.
At higher temperatures, we map out the phase transition lines
at which the gap parameters $\De_1$, $\De_2$ and $\De_3$
describing $ds$-pairing, $us$-pairing and $ud$-pairing respectively,
go to zero in an NJL model.  For small values of $M_s^2/\mu$,
we find that $\De_2$ vanishes first, then $\De_1$, then $\De_3$.
We find agreement with a previous Ginzburg-Landau analysis of
the form of these transitions and find quantitative
agreement with results obtained in full QCD at asymptotic density
for ratios of coefficients in the Ginzburg-Landau potential.
At larger $M_s^2/\mu$,
we find that $\De_1$ vanishes first, then $\De_2$, then $\De_3$.
Hence, we find a ``doubly critical'' point in the  
$(M_s^2/\mu,T)$-plane at which two lines of second order phase
transitions ($\De_1\rightarrow 0$ and $\De_2\rightarrow 0$) 
cross.  Because we do not make any small-$M_s$ approximation,
if we choose a relatively strong coupling leading to large gap
parameters, we are able to pursue the analysis of
the phase diagram all the way up to such large 
values of $M_s$ that there are no strange quarks present. 
\end{abstract}


\maketitle


\section{Introduction}

Our understanding of the properties of matter at high baryon 
number density has developed rapidly in the last several years.
At any densities that are high enough that nucleons are 
crushed into quark matter, 
the quark matter that results must, at sufficiently
low temperatures, be in 
one of a family of color superconducting phases~\cite{reviews}.
The essence of color superconductivity is quark pairing,
driven by the BCS mechanism which operates whenever there
is an attractive interaction between fermions at a Fermi surface.
The QCD quark-quark interaction is strong and is attractive
between quarks that are antisymmetric in color, so we expect 
cold dense quark matter to generically exhibit color superconductivity.
Color superconducting quark matter may well occur in the
cores of compact stars.

It is by now well-established that at asymptotic densities, 
where the up, down and strange quarks can be treated
on an equal footing and the disruptive effects of the
strange quark mass can be neglected, quark matter
is in the color-flavor locked (CFL) phase, in which
quarks of all three colors and all three flavors form
Cooper pairs~\cite{Alford:1998mk}.  The CFL phase is a color superconductor
but is an electromagnetic insulator, with zero electron density.

To describe quark matter as may exist in the cores of 
compact stars, we need consider quark chemical potentials $\mu$
of order 500~MeV at most, meaning that the strange quark
mass $M_s$ must be included: it is expected to be density dependent,
lying between the current mass $\sim 100$~MeV and the
vacuum constituent quark mass $\sim 500$~MeV.  In bulk
matter, as is relevant for compact stars where we
are interested in kilometer-scale volumes, we must
furthermore require electromagnetic and color
neutrality~\cite{Iida:2000ha,Alford:2002kj} (possibly via mixing
of oppositely charged phases)
and allow for equilibration under the weak
interactions. All these factors work to pull apart the
Fermi momenta of the cross-species pairing that 
characterizes color-flavor locking.  At the highest
densities, we expect CFL pairing, but as the density
decreases the combination of nonzero $M_s$ and 
the constraints of neutrality put greater and
greater stress on cross-species pairing, and reduce
the excitation energies of those fermionic quasiparticles
whose excitation would serve to ease the stress by
breaking pairs~\cite{Alford:2003fq}.

If we imagine beginning in the CFL
phase at asymptotic density, reducing the density, and
assume that CFL pairing is disrupted by the heaviness of
the strange quark before color superconducting quark matter
is superseded by the hadronic phase, the next phase down
in density is the gapless CFL phase~\cite{Alford:2003fq,Alford:2004hz}.  
In this phase,
quarks of all three colors and all three flavors still form
Cooper pairs, but there are regions of momentum space in
which certain quarks do not succeed in pairing, and these
regions are bounded by momenta at which certain 
fermionic quasiparticles are gapless.  The material is
an electromagnetic conductor, with a nonzero electron density.

Our goal in this paper is to map the phase diagram of neutral
dense quark matter at nonzero temperature, answering the
question of what phases and phase transitions result
when CFL or gCFL quark matter is heated.  In the case
of CFL quark matter with strange quark mass $M_s=0$, this
question has been answered in Ref.~\cite{Schmitt:2002sc}.  In this most
symmetric setting, there is a single phase transition 
at a temperature $T_c^{\rm CFL}$ below which there is
CFL pairing and above which there is no pairing.  In
mean field theory, 
\begin{equation}
T_c^{\rm CFL}=2^{\frac{1}{3}}\frac{e^\gamma}{\pi}\De_0\ ,
\label{eq:CFLTc}
\end{equation}
where $\De_0$ is the CFL gap parameter at $T=0$, estimated
to be of order 10 to 100~MeV~\cite{reviews}.   The 
enhancement of $T_c/\De_0$ by a factor of $2^{1/3}$ over the standard
BCS value (which our results confirm)   
originates in the fact that in the CFL phase 
with $M_s=0$ there are eight fermionic quasiparticles 
with gap $\De_0$ and one with gap $2\De_0$~\cite{Schmitt:2002sc}.
As at any phase transition at which a superconducting
order parameter melts, 
gauge field fluctuations that are neglected in mean field
theory can elevate $T_c$ and render
the transition first order~\cite{Halperin:1973jh}.  Because the
gauge coupling in QCD is strong, these effects are 
significant~\cite{Matsuura:2003md,Giannakis:2004xt}. They have been 
evaluated to date only at $M_s=0$~\cite{Matsuura:2003md,Giannakis:2004xt}. 
Once $M_s\neq 0$, the mean field analysis alone becomes 
rather involved and we shall therefore leave the inclusion
of fluctuations to future work.

To see why $M_s\neq 0$ results in an intricate 
phase diagram at nonzero temperature, we must introduce
the pairing ansatz that we shall use throughout 
this paper~\cite{Alford:1998mk,Steiner:2002gx,Alford:2003fq,Alford:2004hz}:
\begin{equation}
\langle \psi^\alpha_a C\gamma_5 \psi^\beta_b \rangle \sim 
\Delta_1 \eps^{\alpha\beta 1}\eps_{ab1} \!+\! 
\Delta_2 \eps^{\alpha\beta 2}\eps_{ab2} \!+\! 
\Delta_3 \eps^{\alpha\beta 3}\eps_{ab3}
\label{eq:diquark}
\end{equation} 
Here $\psi^\alpha_a$ is a quark of color $\alpha=(r,g,b)$ 
and flavor $a=(u,d,s)$;
the condensate is a Lorentz scalar, antisymmetric in Dirac indices,
antisymmetric in color 
(the channel with the strongest
attraction between quarks), and consequently
antisymmetric in flavor. 
The gap parameters
$\De_1$, $\De_2$ and $\De_3$ describe down-strange,
up-strange and up-down Cooper pairs, respectively.
At $T=0$, there is 
an insulator-metal transition at $M_s^2/\mu \simeq 2 \De_1$,
at which the CFL insulator with $\De_3\simeq\De_2=\De_1$ 
is replaced by the gapless CFL metal, which has
$\De_3>\De_2>\De_1>0$~\cite{Alford:2003fq,Alford:2004hz}.  
(If the CFL phase is augmented
by a $K^0$-condensate~\cite{Bedaque:2001je}, 
the CFL$\rightarrow$gCFL transition 
is delayed to $M_s^2/\mu\simeq 8\De_1/3$~\cite{Kryjevski:2004jw}.)
An analogous zero temperature metal insulator transition
has been analyzed in Ref.~\cite{Liu:2004mh}.

At $M_s=0$, $\De_1=\De_2=\De_3=\De_{\rm CFL}$,
and $\De_{\rm CFL}$ decreases from $\De_0$ at $T=0$
to 0 at $T_c^{\rm CFL}$.   As soon as $M_s\neq 0$, however,
we can expect that $\De_1$, $\De_2$ and $\De_3$ 
do not all vanish at the same temperature.  This expectation
is evident for the gCFL phase, which has 
$\De_3>\De_2>\De_1>0$ already at $T=0$. We shall see that
it also applies
to the CFL phase with $M_s \neq 0$.  

\begin{figure*}
\includegraphics[width=12cm]{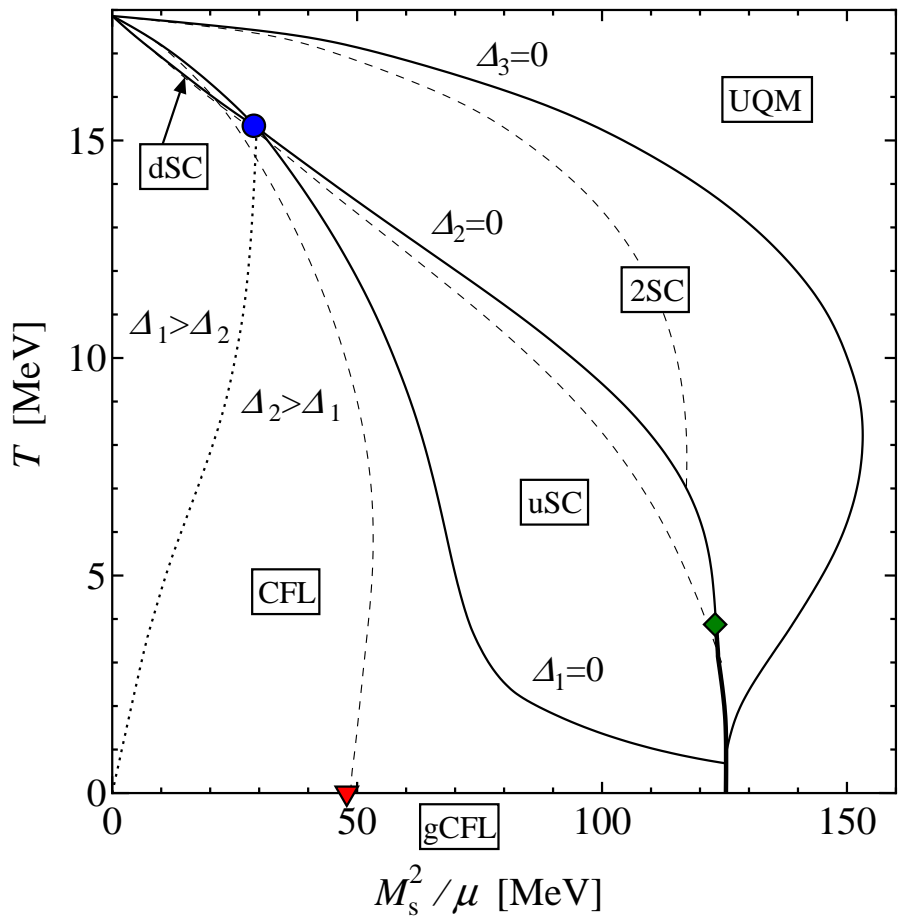}
\caption{Phase diagram of dense neutral quark matter in the 
$(M_s^2/\mu,T)$ plane. Only the solid curves represent
phase transitions. The thin solid curves denote mean field second
order phase transitions at which $\De_1$ or $\De_2$ or $\De_3$ 
vanish.  Two of these lines cross at the circle, which
we call a ``doubly critical'' point.
The heavy solid line is a first order phase transition
that ends at a tricritical point denoted by the diamond. 
The dashed curves indicate locations where, as $M_s^2/\mu$ 
is increased, new gapless modes appear. At any nonzero temperature,
however, there is no physical distinction between a phase with
truly gapless modes and one in which the same modes have excitation
energies of order $T$.  The dashed curves therefore have physical
significance only where they cross $T=0$. In this phase diagram,
this occurs only at the triangle, which denotes the 
zero temperature CFL$\rightarrow$gCFL transition, which
is a quantum critical point.  The dotted line
separates the $\De_1>\De_2$ and $\De_2>\De_1$ regions. 
This phase
diagram is drawn for $\mu=500$~MeV, with $M_s$ and $T$ varying,
and with
the quark-quark interaction strength fixed and
chosen such that the CFL gap is $\De_0=25$~MeV
at $T=0$. We shall describe the model and approximations within which
this phase diagram has been obtained in Section~\ref{sec:model}, and describe
the results summarized by this diagram in Sections~\ref{sec:zero} and \ref{sec:nonzero}.
}
\label{fig:phase_diagram}
\end{figure*}

The distinction between an insulator and a metal is sharp
only at $T=0$. At any nonzero temperature, there will be
some nonzero density of thermally excited fermionic quasiparticles,
some of which are charged.  This means that the CFL$\rightarrow$gCFL
``transition'' should be a crossover at any nonzero 
temperature~\cite{Ruster:2004eg}.  The CFL phase at $T=0$
with $M_s\neq 0$ has fermionic quasiparticles with opposite
charges whose excitation energies differ. This means that
upon heating this phase, the chemical potentials needed
to maintain neutrality are not the same as at zero temperature.
Adjusting the chemical potentials feeds back into the gap equations
for $\De_1$, $\De_2$ and $\De_3$ differently, and these gap parameters
can therefore have different $T$-dependence.  If we start
in the CFL phase at $T=0$, heat, then increase $M_s^2/\mu$ above
that for the zero temperature CFL$\rightarrow$gCFL transition, and
then cool back to $T=0$ we can go from CFL to gCFL without ever
crossing a phase boundary.  We illustrate this by showing
an example phase diagram in Fig.~\ref{fig:phase_diagram}.

The purpose of this paper is to derive and understand the phase
diagram of Fig.~\ref{fig:phase_diagram}. We shall present the
model and approximations that we use in Section~\ref{sec:model}. 
We follow the conventions for naming 
phases used in
previous literature:
\begin{align*}
 & \Delta_1, \Delta_2, \Delta_3 \neq 0 && \text{(g)CFL} \\
 & \Delta_1 = 0, \quad \Delta_2, \Delta_3 \neq 0 && \text{uSC} \\
 & \Delta_2 = 0, \quad \Delta_1, \Delta_3 \neq 0 && \text{dSC} \\
 & \Delta_1 = \Delta_2 = 0, \quad \Delta_3 \neq 0 && \text{(g)2SC} \\
 & \De_1=\De_2=\De_3=0 && \text{UQM,}
\end{align*}
with ``UQM'' meaning unpaired quark matter.
The origin of the remaining names is that in the 2SC phase, only quarks
of two flavors and two colors pair, whereas in the uSC (or dSC)
phase, all $u$ (or all $d$) quarks pair~\cite{Iida:2003cc}.   
The phase diagram features
three lines denoting second order phase transitions 
at which $\De_1$, $\De_2$ or $\De_3$ vanish.
At $M_s=0$, all three gaps vanish at the same temperature but
at generic values of $M_s$,
there are three distinct transition temperatures.
Two of these lines cross at a point denoted by the
circle: at this ``doubly critical point'', $\De_1$ and $\De_2$ 
vanish at the  same temperature.  
The phase
diagram is intricate, and one natural question is to
what extent its features are generic. We address this by 
varying parameters, as we shall discuss in 
Sections~\ref{sec:zero} and \ref{sec:nonzero}.
We shall find that the physics at large $M_s^2/\mu$, where
we find a first order phase transition in Fig.~\ref{fig:phase_diagram},
is not generic, changing qualitatively for larger values
of $\De_0$.  
We find that the shape of the 
second order phase boundaries separating
the CFL, dSC, uSC and 2SC phases 
is generic, as is the existence of the doubly critical point.

Physics in the vicinity of any of these phase transitions
can be analyzed using a Ginzburg-Landau approximation, in
which the relevant $\De$ or $\De$'s are taken to be small.
This analysis has been performed at $M_s=0$ in 
Refs.~\cite{Iida:2000ha,Matsuura:2003md,Iida:2002ev}
and at small but nonzero $M_s$ in Ref.~\cite{Iida:2003cc}.  We shall show
in Section~\ref{sec:GL} that our numerical results for the 
three phase transition lines at small $M_s^2/\mu\De_0$
are in quantitative
agreement with the 
Ginzburg-Landau approximation, and shall show that the 
ratios of coefficients
in the Ginzburg-Landau potential that we obtain in our model
agree quite well with those obtained in full QCD at 
higher densities. At small $M_s^2/\mu\De_0$,
the region of the phase
diagram where the Ginzburg-Landau analysis applies near $T_c$,
$\De_2$ vanishes at a lower critical temperature
than $\De_1$.  In heating gCFL quark matter, however, we
expect $\De_1$ to vanish first because it is already much
smaller at $T=0$.  Our model analysis
can be extended to larger $M_s$ and smaller $T$ 
than the Ginzburg-Landau analysis, allowing us to see how 
the phase diagram fits together, consistent with this expectation.

A phase diagram similar to ours was obtained previously
in Ref.~\cite{Ruster:2004eg}, but our results differ in a crucial, 
qualitative respect at low $M_s^2/\mu$, where
we believe that our results are robust: the authors of
Ref.~\cite{Ruster:2004eg} found that $\De_1$ always
vanishes at a lower temperature than $\De_2$, meaning
that there is no dSC region on their phase diagram.
This also 
disagrees with the Ginzburg-Landau analysis~\footnote{We have learned
from them in private communications that the authors
of Ref.~\cite{Ruster:2004eg} now find an ordering of phase
transitions in agreement with our results and the Ginzburg-Landau
analysis.}.
We shall
detail our approximations in Section~\ref{sec:model}, but we note already
here that whereas in
Refs.~\cite{Alford:2003fq,Alford:2004hz,Ruster:2004eg} $M_s$ was assumed to be
much smaller than $\mu$, 
we do not make such an approximation.  This likely explains the
differences between our results and those of Ref.~\cite{Ruster:2004eg}
at large $M_s^2/\mu$.  

Although not associated with a phase transition, it is interesting
to ask {\it how} the CFL phase ceases to be an insulator as
it is heated. This must happen, given the phase diagram
that we and the authors of Ref.~\cite{Ruster:2004eg} find, but how?  
At small $T$, the number density of
charged quark quasiparticles in the CFL phase is exponentially
small. Because of the interplay between $M_s\neq 0$ and
the constraints of neutrality, the excitation energies of
oppositely charged quasiparticles are not the same and
these thermally excited quasiparticles have a net charge
which must be balanced by a nonzero electron density of
order $\mu_e^3$, where $\mu_e$ is the electron chemical potential.  
Because the quasiparticle densities are proportional to 
the quark Fermi surface area $\sim\mu^2$, where $\mu\gg\mu_e$,
we show that $\mu_e$ ceases to be exponentially suppressed -- rising
rapidly to $\mu_e\sim M_s^2/4\mu$ -- in an insulator-metal crossover
that occurs in a narrow, and quite low, range of temperatures.
We describe this crossover analytically in Section~\ref{sec:crossover}.
There are other charged excitations in the CFL phase, namely
the pseudo-Nambu-Goldstone bosons, that are excited when the system
is heated. We argue quantitatively that their effects are small.

In Section~\ref{sec:model} we detail our model and approximations, stressing
also those approximations used in previous work that we have
not made.  In Section~\ref{sec:zero}, we review results at $T=0$.
In Section~\ref{sec:nonzero}, we present our results at $T\neq 0$, 
analyzing the
phase diagram in Fig.~\ref{fig:phase_diagram} and that for
other values of $\De_0$. In Section~\ref{sec:crossover},
we analyze the insulator-metal crossover that occurs
when the CFL phase is heated.  In Section~\ref{sec:GL}, we make the 
connection between our work and the Ginzburg-Landau analysis
quantitative. We conclude in Section~\ref{sec:conclude}.

\section{Model and Approximations}
\label{sec:model}

\subsection{Pairing ansatz for gap parameters}

We assume that the
predominant diquark condensate is a Lorentz scalar (antisymmetric in
Dirac indices), 
is antisymmetric in color (as is favored energetically), 
and consequently
is antisymmetric in flavor. That is, we assume a condensate
of the form (\ref{eq:diquark}).  
The gap parameters
$\De_1$, $\De_2$ and $\De_3$ 
describe a $9\times 9$ matrix in color-flavor
space that, in the basis $(ru, gd, bs, rd, gu, rs, bu, gs, bd)$,
takes the form
\beq
\label{blocks}
\bm{\Delta}= 
\gamma_5 \cdot 
\newcommand{\mDe}{\makebox[1.2em][r]{$-\De$}} 
\left(
\begin{array}{ccccccccc}
 0 & \Delta_3 & \Delta_2 & 0 & 0 & 0 & 0 & 0 & 0 \\
 \Delta_3 & 0 & \Delta_1 & 0 & 0 & 0 & 0 & 0 & 0 \\
\Delta_2 & \Delta_1 & 0  & 0 & 0 & 0 & 0 & 0 & 0 \\
  0 & 0 & 0 & 0 & \mDe_3 & 0 & 0 & 0 & 0 \\
  0 & 0 & 0 & \mDe_3 & 0 & 0 & 0 & 0 & 0 \\
  0 & 0 & 0 & 0 & 0 & 0 & \mDe_2 & 0 & 0 \\
  0 & 0 & 0 & 0 & 0 & \mDe_2 & 0 & 0 & 0 \\
  0 & 0 & 0 & 0 & 0 & 0 & 0 & 0 & \mDe_1 \\
  0 & 0 & 0 & 0 & 0 & 0 & 0 & \mDe_1 & 0 \\
\end{array}
\right)
\eeq
We see that $(gs,bd)$,
$(rs,bu)$ and $(rd,gu)$
quarks pair with gap parameters $\De_1$, $\De_2$ and $\De_3$
respectively, while the $(ru,gd,bs)$ quarks pair among each other
involving all the $\De$'s. The most important physics that we
are leaving out by making this ansatz is pairing in which
the Cooper pairs are symmetric in color, and therefore also in
flavor.  Diquark condensates of this form break no
new symmetries, and therefore {\em must} arise
in the CFL phase~\cite{Alford:1998mk,Alford:1999pa}. 
However because the QCD interaction
is repulsive between quarks that are symmetric
in color these condensates are numerically
insignificant~\cite{Alford:1998mk,Alford:1999pa,Ruster:2004eg}.

\subsection{CFL symmetries and excitations}
\label{subsec:symmetries}

To set the stage for our analysis, 
and to introduce the excitations that
shall concern us,
we briefly summarize the properties
of the CFL phase~\cite{Alford:1998mk}. 
If we set all three quark masses to zero,
the diquark condensate in the CFL phase (which then
has $\De_1=\De_2=\De_3$) spontaneously breaks the
full symmetry group of QCD, 
\beq
\ba{r@{}l}
 {[SU(3)_{\rm color}]} &\times 
  \underbrace{SU(3)_L \times SU(3)_R}_{\displaystyle\supset [U(1)_Q]}
 \times U(1)_B \\[5ex]
 &\to\quad 
  \underbrace{SU(3)_{C+L+R}}_{\displaystyle\supset [U(1)_{{\tilde Q} }]} 
  \times \mathbb{Z}_2
\ea
\eeq
where $SU(3)_{\rm color}$ and
electromagnetism $U(1)_Q$ are gauged, and the unbroken
$SU(3)_{c+L+R}$ subgroup consists of 
flavor rotations of the left and right quarks 
with equal and opposite color rotations, and contains an unbroken
gauged ``rotated electromagnetism'' 
$U(1)_{\rm \tilde Q}$~\cite{Alford:1998mk,Alford:1999pb}. 
The CFL phase has the largest possible unbroken symmetry
consistent with diquark condensation, achieved by having
all nine quarks participate equally in the
pairing,  and this gives the maximal pairing free energy
benefit. Not surprisingly, {\em ab initio}
calculations valid at asymptotic densities confirm that the
CFL phase is the ground state of QCD in the high density
limit~\cite{Evans:1999at,reviews}. 

The axial flavor $U(1)_A$ rotations are not a symmetry
of QCD, as they are explicitly broken at all densities by
instanton effects.  At asymptotic densities, these effects
become small but at accessible
densities they cannot be 
neglected~\cite{Alford:1998mk,Rapp:1999qa,Schafer:1999fe,Manuel:2000wm,Schafer:2002ty},
and they shall play 
an important role in Section~\ref{subsec:meson}.

In the limit of three massless quarks described above
there are 17 broken symmetry generators in the CFL phase, 8 of which become
longitudinal components of massive gauge bosons and 9 of which
remain as Nambu-Goldstone bosons.  The massive gauge bosons have
excitation energies of order $g\mu$, where the QCD
gauge coupling $g$ is not small, and we neglect them throughout.
In the real world, with its two light quark flavors with
masses $\lesssim 10~{\rm MeV}$, and a medium-weight flavor, the strange 
quark, with mass $\gtrsim 100~{\rm MeV}$,
the eight Nambu-Goldstone bosons
coming from the breaking of chiral symmetries acquire
masses~\cite{Alford:1998mk,Son:1999cm,Hong:1999ei,Manuel:2000wm,Schafer:2002ty,reviews}, 
and furthermore the CFL condensate
may rotate in the $K^0$-direction
within the manifold describing these 
mesons~\cite{Bedaque:2001je}.  We neglect the possibility 
of $K^0$-condensation throughout. We shall see in 
Section~\ref{subsec:meson} that this corresponds to
assuming that the instanton contribution to the $K^0$ mass
is comparable to $\De_0$~\cite{Schafer:2002ty,Reddy:2002xc}, 
which is likely the case for all
but the largest $\De_0$ that we consider. We must also consider
the thermally excited pseudo-Nambu-Goldstone bosons, but
we show in Section~\ref{subsec:meson} that their effects are negligible.
The ninth Nambu-Goldstone boson,
corresponding
to the spontaneous breaking of $U(1)_B$ and hence to 
superfluidity, remains massless even once quark masses
are taken into account and therefore plays a crucial role in many low
energy properties of the CFL phase, for example in its
specific heat, thermal conductivity, neutrino opacity, and
neutrino emissivity at low 
temperatures~\cite{Reddy:2002xc,Jaikumar:2002vg,Kundu:2004mz}.  
However, because it is neutral it will play no role in our analysis.

The excitations of the CFL phase that are most relevant to
our considerations are the nine fermionic quasiparticles,
whose excitation requires the breaking of pairs.
In analyzing the response of the CFL phase to the strange
quark mass at zero temperature, the quasiquark dispersion relations
signal the
instability corresponding to the disruption of pairing 
that constitutes the CFL$\rightarrow$gCFL 
transition~\cite{Alford:2003fq,Alford:2004hz}.
These dispersion relations are even more crucial at nonzero temperature,
because they determine the number densities of thermally
excited quasiparticles, some of which are charged and which
we shall see therefore play a crucial role in the neutrality conditions.
Furthermore, these dispersion relations define the gap
equations that we shall solve at nonzero temperature, again as we
shall see.


\subsection{Color and electric neutrality}

Stable bulk matter must be neutral under
all gauged charges, whether they are spontaneously broken or not.
Otherwise, the net charge density would create large electric fields,
making the energy non-extensive.
In the case of the electromagnetic gauge symmetry, this simply 
requires zero charge density.
In the case of the color gauge symmetry, the formal requirement is
that a chunk of quark matter should be a color singlet, i.e.,
its wavefunction should be invariant under a general color gauge
transformation. Color neutrality, meaning equality
in the numbers of red, green
and blue quarks, 
is a less stringent constraint.
A color singlet state is also
color neutral, whereas the opposite is not necessarily true. 
However it has
been shown that 
the projection of a color neutral state
onto a color singlet state costs no extra free energy in the
thermodynamic limit~\cite{Amore:2001uf}.  Analyzing the consequences of the
requirement of color
neutrality therefore suffices for our purposes.

In nature, electric and
color neutrality are enforced by the dynamics of the electromagnetic
and QCD gauge fields, whose zeroth components serve as chemical potentials 
which take on values that enforce 
neutrality~\cite{Alford:2002kj,Gerhold:2003js}.  
Since we are
limiting ourselves to color neutrality and not color singletness we
have to consider only the $U(1)\times U(1)$ 
diagonal subgroup of the color
gauge group. This subgroup is generated by the 
diagonal generators 
$T_3=\diag(\half,-\half,0)$ and $T_8=\diag(\third,\third,-\twothirds)$
of the $SU(3)$ gauge group. Electromagnetism is generated by
$Q=\diag(\frac{2}{3},-\frac{1}{3},-\frac{1}{3})$ in flavor space ($u$, $d$, $s$). 
The zeroth components of the respective gauge fields serve as
chemical
potentials $\mu_3$ and $\mu_8$ coupled to $T_3$ and $T_8$ charges,
and as an electrostatic potential $\mu_e$ coupled to the
{\em negative} electric charge $Q$.  (We make this last choice
so that $\mu_e>0$ corresponds to a density of electrons, not
positrons.) The dynamics of the 
gauge potentials then require that the charge
densities, which are the derivatives of the free energy with respect to
the chemical potentials, must vanish:
\beq
\label{neutrality}
\ba{rcl}
n_Q = \phantom{-}\dsp\frac{\partial \Omega}{\partial\mu_e}&=& 0 \\[2ex]
n_3= -\dsp\frac{\partial \Omega}{\partial\mu_3}&=& 0 \\[2ex]
n_8= -\dsp\frac{\partial \Omega}{\partial\mu_8}&=& 0\ .
\ea
\eeq
In an NJL
model with fermions but no gauge fields  that we shall
employ, one has to
introduce the chemical potentials $\mu_e$, $\mu_3$ and
$\mu_8$ ``by hand'' in order to enforce 
color and electric neutrality in the same way that gauge field
dynamics does in QCD~\cite{Alford:2002kj}.

A generic diquark condensate will be neither electrically nor
color neutral, so it will spontaneously break these gauge symmetries.
However it may be neutral under a linear combination
of electromagnetism and color. Indeed, any condensate of
the form (\ref{eq:diquark}) is neutral with respect to
the ``rotated electromagnetism'' generated by 
$\tilde Q = Q -T_3 -\half T_8$, so $U(1)_{\tilde Q}$ is never broken.
This means that the corresponding gauge boson
(the ``$\Qtilde$-photon''), a mixture of the ordinary photon and
one of the gluons, remains massless. 
In both the CFL and gCFL phases,
the rest of the
$SU(3)_{\rm color} \times U(1)_Q$
gauge group is spontaneously broken, meaning
that the combination of the photon and gluons
orthogonal to the $\Qtilde$-photon, and all the
other gluons, become massive by the Higgs mechanism.

In the CFL phase at $T=0$, many consequences of 
the neutrality equations can be analyzed
in a model-independent way~\cite{Alford:2002kj}. 
A central result is that the free energy $\Omega$ is independent
of 
\begin{equation}
 \mu_{\Qtilde} = -{\txt \frac{4}{9}}\bigl(\mue
  +\mu_3 + \half\mu_8\bigr)\ ,
\label{muQtilde}
\end{equation}
the chemical potential for $\tilde Q$-charge, as long as 
$\mu_{\Qtilde}$ is not so large that it disrupts CFL pairing.
This means that the CFL phase is a $\Qtilde$-insulator 
at $T=0$~\cite{Rajagopal:2000ff}.
Translating back into $\mue$, $\mu_3$ and $\mu_8$,
this means that two of these three can be fixed as functions of the third:
\begin{align}
 \mu_3 &= \mue,
\label{eq:CFL_mu3} \\
 \mu_8 &= -\frac{\ms^2}{2\mu}+\frac{\mue}{2},
\label{eq:CFL_mu8}
\end{align}
where $\mue$ can lie anywhere within the range
$-\Delta_2+\frac{M_s^2}{2\mu}<\mu_e<+\Delta_3$.
If the lower (upper) bound of this range is violated 
$rs$-$bu$ ($rd$-$gu$) pairing is disrupted: this range
can be thought of as the bandgap for the CFL insulator~\cite{Alford:2003fq}.
The free energy $\Omega$ has a plateau within this
range of $\mu_\Qtilde$. 
Electrons add a new term $\sim \mu_e^4$ to $\Omega$, adding
a gentle curvature to the plateau, and 
selecting $\mu_e=0$, meaning zero electron
density, as the correct solution for neutral CFL
matter at $T=0$.  ($\mu_3$ and $\mu_8$ are then given by
(\ref{eq:CFL_mu3},\ref{eq:CFL_mu8}) with $\mu_e=0$.)
At finite
temperature where 
quasi-quarks with non-trivial $\Qtilde$ are excited,
we shall see that because $\Omega$ has an only-very-gently-curved
plateau, it will not take much thermal excitation 
of quasiquarks to shift
$\mu_e$ significantly away from zero.

\subsection{A Model for the Thermodynamic Potential}

We are interested in physics at non-asymptotic densities, and therefore
cannot use weak-coupling methods.  We are interested in physics at
zero temperature and high density, at which the fermion sign problem
is acute and the current methods of lattice QCD can therefore not be
employed.  For this reason, we need to introduce a model in which the
interaction between quarks is simplified, while still respecting the
symmetries of QCD, and in which the effects of $M_s$, $\mu_e$, $\mu_3$,
$\mu_8$ and $T$ on CFL pairing can all be investigated.  The natural
choice is to model the interactions between quarks using a point-like
four-fermion interaction, which we shall take to have the quantum
numbers of single-gluon exchange.
We work in Euclidean space.
Our partition function
$\cal Z$ and free energy density $\Om$ are then defined by
\beq
\label{pathintegral}
\ba{rcl}
{\cal Z} &=&\dsp\e^{-\beta V \Omega}= {\cal N}\int {\cal D}\psibar{\cal D}\psi
  \exp\Bigl(\int {\cal L}(x)\,d^4x\Bigr) \\[4ex]
{\cal L}(x) &=&\dsp \psibar(i\feyn{\partial} + \feyn{\bm{\mu}} - \bm{M})\psi 
 -\frac{3}{8}G(\psibar\Gamma_\mu^A\psi)(\psibar\Gamma^\mu_A\psi)
\ea
\eeq
where the fields live in a box of volume $V$ and Euclidean time length
$\be=1/T$, and $\feyn{\bm{\mu}}=\bm{\mu}\gamma_4$. 
The interaction vertex has the 
color, flavor, and spin structure
of the QCD gluon-quark coupling, $\Gamma_\mu^A = \gamma_\mu T^A$
with the $T_A$ normalized such that $\Tr \, T_A T_B = 2 \delta_{AB}$. 
The mass matrix $\bm{M}=\diag(0,0,M_s)$ in flavor space.
The chemical potential $\bm{\mu}$ is a diagonal color-flavor 
matrix depending on
$\mu$, $\mu_e$, $\mu_3$ and $\mu_8$. The normalization of the
four-fermion coupling
$3G/8$ is as in the first paper in Ref.~\cite{reviews}.
In real QCD, the ultraviolet modes decouple because of asymptotic freedom
but, in the NJL model, we have to add this feature by hand, through
a UV momentum cutoff $\Lambda$ in the momentum integrals.
The model therefore has two parameters, the four-fermion coupling $G$
and the three-momentum cutoff $\Lambda$,  
but it is more useful to parameterize the interaction
in terms of a physical quantity, namely
the CFL gap parameter at $M_s=T=0$ at a reference chemical potential
that we shall take to be $500$~MeV. We shall call this
reference gap $\De_0$. 
We have checked that if we vary the cutoff $\Lambda$
by 20\% while simultaneously
varying the bare coupling $G$ so as to keep $\De_0$ fixed,
then our results change by a few percent at most.
All the results that we present are for $\Lambda=800$~MeV.
Most of the results that we present are
for a coupling strength chosen such that 
$\De_0=25$~MeV, but we shall also discuss results
obtained with stronger couplings for which $\De_0=40$~MeV
and $\De_0=100$~MeV.

The free energy $\Omega$ obtained
from the Lagrangian (\ref{pathintegral}) upon making
the ansatz (\ref{eq:diquark}) for the diquark
condensate and working in the mean field approximation
has been presented in Ref.~\cite{Alford:2004hz}.
More sophisticated derivations exist in the literature~\cite{reviews},
but since we are 
assuming that the only condensate is of the form 
(\ref{eq:diquark}) we can
Fierz transform the interaction to yield products of
terms that appear in (\ref{eq:diquark}), and discard
all the other terms that arise in the Fierz transformed
interaction which would anyway vanish after making the
mean field approximation.  
This yields
\beq
{\cal L}_{\rm int} = \frac{G}{4} \sum_\eta (\bar\psi P_\eta \bar\psi^T)
(\psi^T \bar P_\eta \psi)\ ,
\label{fierzed}
\eeq
where 
\beq
\left(P_\eta\right)^{\al\be}_{ij} = C\ga_5 \eps^{\al\be \eta} \eps_{ij\eta}
\quad\mbox{(no sum over $\eta$)} 
\label{Pdefinition}
\eeq
and $\bar P_\eta = \ga_4 P^\dag_\eta \ga_4$. 
The index $\eta$ labels the pairing channel:
$\eta=1,2,$ and $3$ correspond to $d$-$s$ pairing,
$u$-$s$ pairing, and  $u$-$d$ pairing.
The gap parameters can then be defined precisely as
\begin{equation}
 \Delta_\eta = \half G
  \langle\psi^{\rm T}\bar{P}_\eta\psi\rangle\ .
\label{eq:gap_def}
\end{equation}
As derived in Ref.~\cite{Alford:2004hz}, the free energy can then
be written as
\begin{align}
 & \Omega = -\frac{1}{4\pi^2}\int_0^\Lambda \!\!dp\,
  p^2 \sum_j \Bigl\{ |\varepsilon_j(p)| \!+\! 2T\ln
  \bigl(1+\mathrm{e}^{-|\varepsilon_j(p)|/T}\bigr)\Bigr\} \notag\\
 & \quad +\frac{1}{G}\bigl(\Delta_1^2 \!+\! \Delta_2^2 \!+\! \Delta_3^2)
   -\frac{\mue^4}{12\pi^2} \!-\! \frac{\mue^2 T^2}{6}
   \!-\! \frac{7\pi^2 T^4}{180}\ ,
\label{eq:potential}
\end{align}
where the electron contribution for
$T\neq 0$ has been included. 
The functions $\varepsilon_j(p)$
are the dispersion relations for the fermionic quasiparticles.
They are not explicitly $T$-dependent, but they do depend
on the $\De$'s and $\mu$'s which are $T$-dependent.
The quasiquark dispersion relations $\varepsilon_j(p)$
are the values of the energy at which the propagator diverges,
\beq
\label{disprel}
\det S^{-1}(-i\varepsilon_j(p),p)=0\ .
\eeq
Here, the inverse propagator $S^{-1}$ is the $72\times 72$
matrix 
\beq
\label{invprop}
S^{-1}(p) = \left( \ba{cc}
\feyn{p} + \feyn{\mu}-\bm{M} & P_\eta\De_\eta \\
\bar P_\eta\De_\eta &  (\feyn{p} - \feyn{\mu}+\bm{M})^T \ea \right)
\eeq
that acts on Nambu-Gor'kov spinors 
\beq
\Psi = \left(\ba{c}\psi(p)\\ \bar\psi^T(-p)\ea\right),\quad
\bar\Psi = \Bigl( \bar\psi(p)\ \psi^T(-p) \Bigr)\ .
\eeq 
The sum in (\ref{eq:potential}) is understood to run over 36 roots,
because the Nambu-Gor'kov formalism has artificially made each
$|\varepsilon_j|$ doubly degenerate.  There is a further (physical)
degeneracy coming from spin, meaning that there are only 18 distinct
dispersion relations to be found.  
The free energy
of any solution to the gap equations obtained from this mean field
free energy is the same as that obtained from the 
Cornwall-Jackiw-Tomboulis
effective potential~\cite{Cornwall:vz}, 
with the mean field  approximation implemented via dropping
the contribution of 2PI diagrams.

We shall evaluate $\Omega$ numerically, meaning that at
every value of the momentum $p$ we must find the zeros of the
inverse propagator.  In order to do this numerically, we
follow Ref.~\cite{Steiner:2002gx} and note that it
is equivalent and faster to find the eigenvalues
of the ``Dirac Hamiltonian density'' ${\cal H}(p)$,
which is related to $S^{-1}$ by
\begin{equation}
 \det S^{-1}(p_4,p) = \det\Bigl[\gamma^4\bigl(i p_4 -{\cal H}(p)
  \bigr)\Bigr]\ ,
\end{equation}
which makes it clear that eigenvalues of ${\cal H}(p)$ are
zeros of $S^{-1}$.
Like $S^{-1}$, 
the Dirac Hamiltonian ${\cal H}(p)$ is a $72\times72$ matrix in color,
flavor, spin, and Nambu-Gor'kov space. This matrix can be decomposed
into 4 blocks -- three $16\times16$ matrices for the 
($rd$-$gu$), ($rs$-$bu$) and ($gs$-$bd$)
pairing sectors, and one $24\times24$ matrix for the
($ru$-$gd$-$bs$) sector. The absolute values of the eigenvalues are
quadruply degenerate due to the Nambu-Gor'kov doubling and the spin
degeneracy.

We have found considerably simpler expressions for
the ($rd$-$gu$), ($rs$-$bu$) and ($gs$-$bd$) sectors. 
The calculation becomes
easier if one adopts a different 
choice of Nambu-Gor'kov basis, i.e.
$\Psi=(\psi(p),C\bar{\psi}^{\rm T}(-p))^{\rm T}$ as in
Ref.~\cite{Buballa:2001gj}. Then the Nambu-Gor'kov and spin degeneracies
are manifest, as the $16\times 16$ matrix separates into 
$4\times 4$ blocks with the form
\begin{equation}
 {\cal H}_{\text{($gs$-$bd$)}}=\left( \begin{array}{cccc}
  -\mu_{bd} & p & 0 & i\Delta_1 \\
  p & -\mu_{bd} & -i\Delta_1 & 0 \\
  0 & i\Delta_1 & \mu_{gs}\!+\!\ms & p \\
  -i\Delta_1 & 0 & p & \mu_{gs}\!-\!\ms
 \end{array} \right)
\label{eq:H4by4}
\end{equation}
for the ($gs$-$bd$) sector and likewise for other sectors. 
Each of the 4 eigenvalues of this matrix, and of its
counterparts for the other sectors, 
contributes
twice (for spin) in the sum in (\ref{eq:potential}).
The quark chemical potentials occurring in (\ref{eq:H4by4})
and below are defined straightforwardly:
\beq
\ba{rcl}
 \mu_{ru}&=& \mu-\twothirds\mu_e+\half\mu_3+\third\mu_8,\\[1ex]
 \mu_{gd}&=& \mu+\third\mu_e-\half\mu_3+\third\mu_8,\\[1ex]
 \mu_{bs}&=& 
\mu+\third\mu_e-\twothirds\mu_8,\\[3ex]
\mu_{rd}&=& \mu+\third\mu_e+\half\mu_3+\third\mu_8,\\[1ex]
 \mu_{gu}&=& \mu-\twothirds\mu_e-\half\mu_3+\third\mu_8,\\[3ex]
\mu_{rs}&=& 
\mu+\third\mu_e+\half\mu_3+\third\mu_8,\\[1ex]
 \mu_{bu}&=& \mu-\twothirds\mu_e-\twothirds\mu_8,\\[3ex]
\mu_{gs}&=& 
\mu+\third\mu_e-\half\mu_3+\third\mu_8,\\[1ex]
 \mu_{bd}&=& \mu+\third\mu_e-\twothirds\mu_8\ . \\[1ex]
\ea
\label{mudefinitions}
\eeq

In the ($ru$-$gd$-$bs$) sector, we could not find any simple way to
make the spin degeneracy manifest, but the Nambu-Gor'kov
degeneracy is manifest as the $24\times 24$ matrix separates
into $12\times 12$ blocks with the form
\begin{widetext}
\begin{equation}
 {\cal H}_{\text{($ru$-$gd$-$bs$)}} =\left( \begin{array}{cccccccccccc}
-\mu_{ru} & p & 0 & -i\Delta_3 & 0 & -i\Delta_2 & 0 &
 0 & 0 & 0 & 0 & 0 \\
p & -\mu_{ru} & i\Delta_3 & 0 & i\Delta_2 & 0 & 0 & 0
 & 0 & 0 & 0 & 0 \\
0 & -i\Delta_3 & \mu_{gd} & p & 0 & 0 & 0 & -i\Delta_1
& 0 & 0 & 0 & 0 \\
i\Delta_3 & 0 & p & \mu_{gd} & 0 & 0 & i\Delta_1 & 0 &
0 & 0 & 0 & 0 \\
0 & -i\Delta_2 & 0 & 0 & \mu_{bs}\!+\!\ms & p & 0 & 0 & 0 &
-i\Delta_1 & 0 & 0 \\
i\Delta_2 & 0 & 0 & 0 & p & \mu_{bs}\!-\!\ms & 0 & 0 & i\Delta_1
& 0 & 0 & 0 \\
0 & 0 & 0 & -i\Delta_1 & 0 & 0 & -\mu_{bs}\!+\!\ms & p & 0 &
0 & 0 & -i\Delta_2 \\
0 & 0 & i\Delta_1 & 0 & 0 & 0 & p & -\mu_{bs}\!-\!\ms & 0 &
0 & i\Delta_2 & 0 \\
0 & 0 & 0 & 0 & 0 & -i\Delta_1 & 0 & 0 & -\mu_{gd} & p & 0 &
-i\Delta_3 \\
0 & 0 & 0 & 0 & i\Delta_1 & 0 & 0 & 0 & p & -\mu_{gd} & i\Delta_3
& 0 \\
0 & 0 & 0 & 0 & 0 & 0 & 0 & -i\Delta_2 & 0 & -i\Delta_3 &
\mu_{ru} & p \\
0 & 0 & 0 & 0 & 0 & 0 & i\Delta_2 & 0 & i\Delta_3 & 0 & p
& \mu_{ru}
 \end{array} \right).
\label{eq:12matrix}
\end{equation}
\end{widetext}
The 12 eigenvalues of this matrix 
each contribute once in the sum in (\ref{eq:potential}). 
They occur in degenerate pairs due to spin.

A stable, neutral phase must 
minimize the free energy (\ref{eq:potential}) 
with respect to variation of the three gap parameters 
$\Delta_1$, $\Delta_2$, $\Delta_3$, meaning it must satisfy
\beq
\label{gapeqns}
\dsp\frac{\partial\Omega}{\partial\Delta_1} = 0, \qquad 
\dsp\frac{\partial\Omega}{\partial\Delta_2} = 0, \qquad
\dsp\frac{\partial\Omega}{\partial\Delta_3} = 0\ ,
\eeq
and it must satisfy the three neutrality conditions
\eqn{neutrality}.
The gap equations (\ref{gapeqns}) 
and neutrality equations \eqn{neutrality} 
form a system of six coupled integral equations 
with unknowns the three gap parameters and $\mu_e, \mu_3$ and $\mu_8$.
We have solved these equations
numerically at a grid of values of $T$ 
and $M_s^2/\mu$.
We evaluate $\Omega$ using (\ref{eq:potential}), finding the
$\varepsilon_j$ by determining the eigenvalues of the matrices
specified explicitly above.  We evaluate partial derivatives
of $\Omega$ using finite difference methods. (We used
three and five point finite difference evaluations of derivatives,
with the interval in the $\De$ or $\mu$ with respect to which
$\Omega$ is being differentiated taken as $0.04$~MeV.)
We have worked entirely with $\mu=500$~MeV, 
varying $M_s^2/\mu$ by varying $M_s$,
and have worked
at three values of the coupling $G$ chosen so that 
$\De_0$, the gap parameter
at $M_s=T=0$, takes on the values $\De_0=25$, 40 and 100~MeV.

\subsection{Approximations made and not made}

Having described our calculation in explicit detail,
we close this section by enumerating the approximations that 
we are making, and stressing several that we have not made.

We have made the following approximations:
\begin{enumerate}
\item
We work throughout in a mean field approximation.  Gauge field
fluctuations will surely be important, and must be included
in future work.
\item
We neglect contributions to the condensates that are symmetric in
color and flavor: these are known to be
present and small, smaller 
than the contributions that we include 
by at least
an order of magnitude~\cite{Alford:1998mk,Alford:1999pa,Ruster:2004eg}.
\item
We choose light quark masses $M_u=M_d=0$
and we treat the constituent strange quark mass $M_s$ as a parameter,
rather than solving for an $\langle \bar s s \rangle$ condensate.
These approximations 
should be improved upon, along the lines of 
Refs.~\cite{Steiner:2002gx,Buballa:2001gj}.
In future studies in which the $\langle \bar q q \rangle$ condensates
are solved for dynamically,
the six-quark 't Hooft interaction induced by instantons
must be included, since it introduces terms of the form
$\langle\bar{s}s\rangle|\Delta_3|^2$, for example.
%

\item
We ignore meson condensation in both the CFL phase and gCFL phases.
We shall see in Section~\ref{subsec:meson} that in the CFL phase 
this means that we are assuming that the
instanton contribution to the $K^0$ mass is comparable to
or larger than $\De_0$~\cite{Schafer:2002ty,Reddy:2002xc}.  
Meson condensation in the gCFL phase
has yet to be analyzed.
\end{enumerate}
We expect that these approximations have quantitative effects, but
none preclude a qualitative understanding of the phase diagram.

The authors of Refs.~\cite{Alford:2003fq,Alford:2004hz} made further 
simplifying
assumptions. They worked only to leading nontrivial order in 
$\De_1$, $\De_2$, $\De_3$, $\mu_e$, $\mu_3$ and $\mu_8$.
They neglected the effects of antiparticles. And, most serious,
they incorporated $M_s$ only via its leading effect, namely
as an effective shift $-M_s^2/\mu$ in the chemical potentials
of the strange quarks. This limits the regime of applicability of their
results to $M_s \ll \mu$.  For $\De_0\sim 100 \MeV$, there is pairing
even at $M_s \gtrsim \mu$ and our calculation runs into no
difficulties in this regime. (We shall show some results,
even though this likely corresponds
to such small $\mu$ that in reality the hadronic phase has taken over.)
In the next section, we shall present our $T=0$ results. Those
with $\De_0=25$~MeV, as in Refs.~\cite{Alford:2003fq,Alford:2004hz}, 
are in very good
agreement with the results of Refs.~\cite{Alford:2003fq,Alford:2004hz}. 
This provides
quantitative evidence 
for the validity of
the approximations made in Refs.~\cite{Alford:2003fq,Alford:2004hz} 
at $\De_0=25$~MeV.


\section{Zero temperature results}
\label{sec:zero}

In this section we shall present the results of our model
analysis at $T=0$.   Here and
throughout this paper, we set $\mu=500$~MeV 
corresponding to a baryon density about ten
times that in nuclear matter, which is 
within the range of expectations for the density at the
center of compact stars.  In reality, both $M_s$ and 
$\De_0$ vary with $\mu$, and our knowledge
of both is uncertain. Our model is by construction
well-suited to study the disruptive effects of $M_s$ 
on pairing but it is not adequate to fix the density-dependent 
values of either $M_s$ or $\De_0$ quantitatively. In
this study, therefore, as in 
Refs.~\cite{Alford:2002kj,Alford:2003fq,Alford:2004hz,Ruster:2004eg} 
we treat both 
$M_s$ and $\De_0$ simply as
parameters.  We shall plot our results versus $M_s^2/\mu$,
rather than versus $M_s$ itself, because the strength of the disruptive
effects introduced by the strange quark mass are characterized by
$M_s^2/\mu$. For example, the splitting between Fermi surfaces
in unpaired quark matter is of order $M_s^2/4\mu$, and the 
CFL$\rightarrow$gCFL transition occurs where $M_s^2/\mu\simeq 2\De_1$.
Although we only present results at a single value of $\mu$, 
because the effects of $M_s$ are controlled by $M_s^2/\mu$,
and because in reality 
$M_s$ is itself a decreasing function of $\mu$,  one
can think of our plots as giving a qualitative description
of the effects of varying $\mu$, with high density at 
small $M_s^2/\mu$ and low density at large $M_s^2/\mu$.
This description is only qualitative because our plots
are each made with some fixed value of $\De_0$, whereas
in reality $\De_0$ depends on $\mu$.

\begin{figure}
\includegraphics[width=8cm]{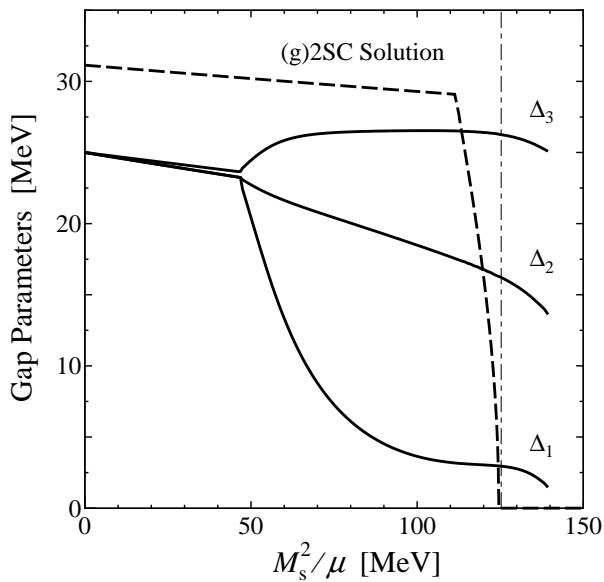}
\caption{Gap parameters $\Delta_1$, $\Delta_2$ and $\Delta_3$ as a
function of $\mssq$ at $T=0$.  The solid curves show the
CFL/gCFL solution, with the CFL$\rightarrow$gCFL transition
occurring where $\mssq \simeq 2 \De_1$.  There is a first order
phase transition between the gCFL phase and unpaired quark
matter at $\mssq=125.3\MeV$, denoted by the
the thin vertical line. To the right of this line,
the gCFL solution is metastable. We also find a neutral
2SC solution, with $\De_3$ given by the dashed curve in the figure,
which undergoes a transition to the gapless 2SC
phase of Refs.~\cite{Shovkovy:2003uu} at $\mssq=112\MeV$.
However, from  Fig.~\ref{fig:energy_zero} we see that
the (g)2SC solution is everywhere metastable, having
a larger free energy than the (g)CFL solution at the same $\mssq$.
}
\label{fig:gaps_zero}
\end{figure}

\begin{figure}
\includegraphics[width=8cm]{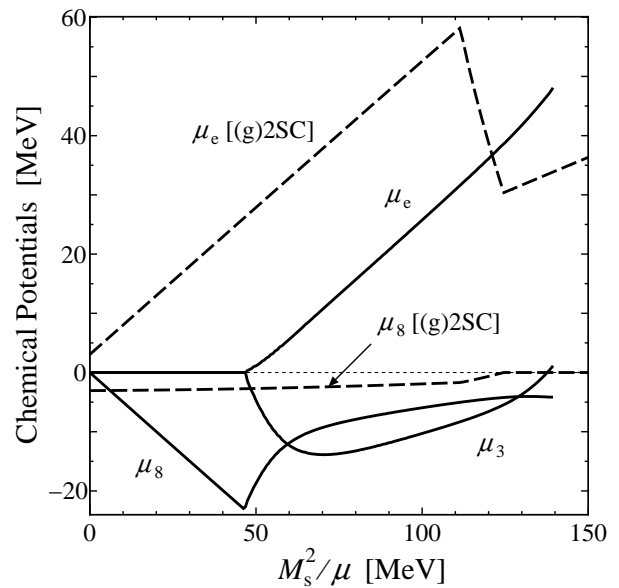}
\caption{Chemical potentials $\mue$, $\mu_3$ and $\mu_8$ as a
function of $\mssq$ at $T=0$, for the (g)CFL and (g)2SC solutions
of Fig.~\ref{fig:gaps_zero}.  ($\mu_3=0$ in the (g)2SC solution
and so is not shown.) Beyond the $\mssq$ at which 
the g2SC solution ends, the dashed curve shows the chemical
potentials ($\mu_e\neq 0$  and $\mu_8=0$) for 
neutral unpaired quark matter.}
\label{fig:chem_zero}
\end{figure}

\begin{figure}
\includegraphics[width=8cm]{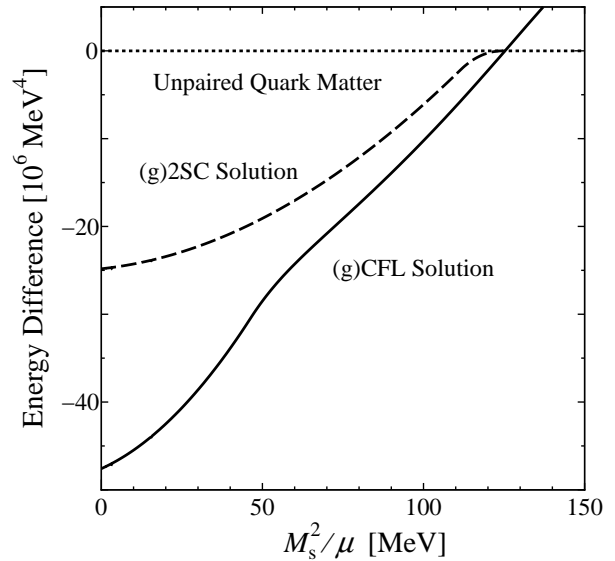}
\caption{Free energies of the (g)CFL and (g)2SC solutions
of Fig.~\ref{fig:gaps_zero} at $T=0$, relative to that of neutral unpaired
quark matter. 
}
\label{fig:energy_zero}
\end{figure}

We begin by choosing the four fermion coupling
$G$ so that $\De_0=25$~MeV, as in Refs.~\cite{Alford:2003fq,Alford:2004hz}.
The solid curves in 
Fig.~\ref{fig:gaps_zero} show the gap parameters as a function of
$\mssq$, and those in Fig.~\ref{fig:chem_zero}
show the chemical potentials. These plots are in very good
agreement with those of Refs.~\cite{Alford:2003fq,Alford:2004hz}, 
indicating that
the approximations made in that paper that we have
dispensed with here, chiefly the small-$M_s^2/\mu^2$ assumption,
were good approximations for $\De_0=25$~MeV.
For small $\mssq$, we see the CFL phase
with $\De_1=\De_2\simeq \De_3$.  The small difference
between $\De_1=\De_2$ and $\De_3$, less
than 2\% everywhere within the CFL
phase, is an example of an effect that we can see but that
cannot be seen in the small-$M_s^2/\mu^2$ approximation in which
$M_s$ is approximated as a shift in the chemical potential
for the strange quarks~\cite{Alford:2003fq,Alford:2004hz}. 
The CFL$\rightarrow$gCFL
transition occurs at $\mssq=46.8\MeV$, which is very close
to $2\De_1$, since $\De_1=23.2\MeV$ at $\mssq=46.8\MeV$.
We see from Fig.~\ref{fig:chem_zero} that
$\mu_e\neq 0$ in the gCFL phase, indicating a nonzero
electron density.
As discussed in detail in Refs.~\cite{Alford:2003fq,Alford:2004hz}, 
the negative 
$\Qtilde$-charge of the electrons is balanced by that
of unpaired $bu$ quarks, which have $\Qtilde=+1$, occurring
in a narrow shell in momentum space.  There is a larger
shell in momentum space, whose width grows with increasing
$\mssq$ in the gCFL phase, within which there are unpaired $bd$-quarks.
This ``blocking region''
of momentum space does not contribute in the $\De_1$
gap equation, and $\De_1$ is consequently
driven down.  The gap equations
and neutrality conditions are all coupled, and the consequences
of the reduction in $\De_1$ and the increase in $\mu_e$ are manifest
in all the curves in Figs.~\ref{fig:gaps_zero}~and~\ref{fig:chem_zero}.

As $\mssq$ increases further, the gCFL solution eventually ceases
to exist at $\mssq=139\MeV$. The gCFL solution 
to the gap equations is a minimum of $\Omega$ with
respect to variation of the $\De$'s for  $\mssq<139\MeV$,
becomes an inflection point at $\mssq=139\MeV$, and for
larger $\mssq$ there is no such solution. This is analyzed
in greater detail in Refs.~\cite{Alford:2003fq,Alford:2004hz}.
The fact that the gCFL solution disappears
indicates that there should be some other minimum with
lower free energy, and indeed as shown in Fig.~\ref{fig:energy_zero}
we find that a first-order phase transition at which the 
gCFL free energy crosses above that of unpaired quark matter 
has occurred at  $\mssq=125\MeV$, indicated in Fig.~\ref{fig:gaps_zero}
by the vertical line.  In Ref.~\cite{Alford:2004hz}, the first order
phase transition and the termination of the gCFL phase at
a point of inflection of the free energy occur at $\mssq=130\MeV$
and $\mssq=144 \MeV$ respectively. Therefore, the errors in these
quantities introduced by the small-$M_s^2/\mu^2$ approximation,
used in Refs.~\cite{Alford:2003fq,Alford:2004hz} but not here, are 
about 4\% at these values of $\mssq$.

As in Ref.~\cite{Alford:2004hz}, we find an additional 
neutral 2SC solution, whose gap parameter $\De_3$
and free energy are shown in 
Figs.~\ref{fig:gaps_zero}~and~\ref{fig:energy_zero}.
At $M_s=0$, the 2SC gap is $2^{1/3}\De_0$~\cite{Schafer:1999fe,reviews}.
For $\mssq$ below the 
2SC$\rightarrow$g2SC transition at $\mssq=112\MeV$, 
$rd$-$gu$ and $ru$-$gd$ pairing occur at all momenta;
above this transition, in the gapless 2SC 
phase~\cite{Shovkovy:2003uu,Gubankova:2003uj}, there is a 
blocking region~\cite{Alford:2000ze} 
in momentum space in which one finds unpaired $rd$ and $gd$ quarks,
and $\De_3$ drops precipitously. An analogue
of the g2SC phase~\cite{Sarma} (and in fact an analogue
of the gCFL phase~\cite{Alford:1999xc})
were first analyzed in contexts in which they were
metastable, but it was shown in Ref.~\cite{Shovkovy:2003uu}
that the g2SC phase could be stabilized in two-flavor
quark matter by the constraints of neutrality.
However, we see in Fig.~\ref{fig:energy_zero} that
in this three-flavor quark matter setting,
the (g)2SC solution everywhere has a larger free
energy than the (g)CFL solution at the same $\mssq$, and
is therefore metastable. The value of $\mssq$ at which
$\De_3\rightarrow 0$ in the g2SC solution is less than
1~MeV below the $\mssq$
at which the gCFL phase becomes metastable.  In contrast,
in Ref.~\cite{Alford:2004hz} the g2SC solution persists to an
$\mssq$ that is less than 1~MeV above 
that at which the gCFL phase
free energy crosses that of unpaired quark matter.
This is the one instance where the small-$M_s^2/\mu^2$
approximation made in Ref.~\cite{Alford:2004hz} leads one (slightly) astray,
as it predicts a (very narrow) $\mssq$-window in which the g2SC phase
is favored and we find no such window.  
However, we shall see below that the physics at 
values of $\mssq$ that are this large compared to $\De_0$ 
is anyway not robust, changing qualitatively with increasing $\De_0$.

\begin{figure}
\includegraphics[width=8cm]{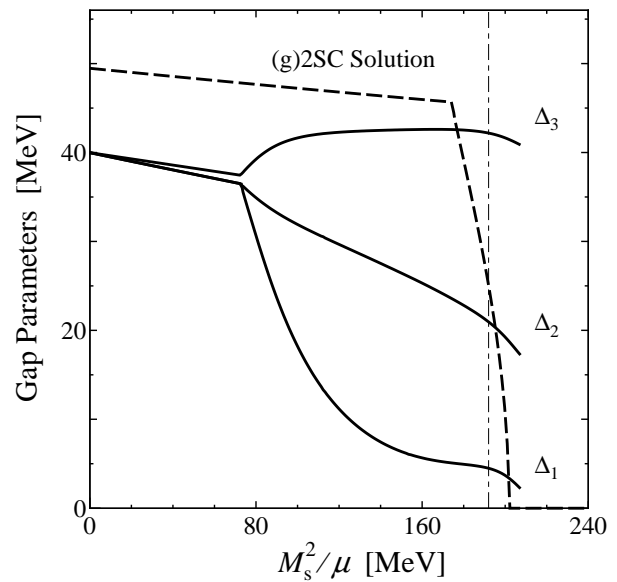}
\caption{Gap parameters for a stronger interaction than
that in Fig.~\ref{fig:gaps_zero}, chosen such that
$\Delta_0=40\MeV$ for $\ms=0$. As in Fig.~\ref{fig:gaps_zero},
the CFL$\rightarrow$gCFL transition occurs where $\mssq\simeq 2\De_1$.
Here, however, the gCFL
phase becomes metastable (at the thin vertical line) at 
a value of $\mssq$ above which
there is a g2SC solution.  There is a first
order gCFL$\rightarrow$g2SC transition at the thin vertical
line, followed at a larger $\mssq$ by a second order transition
at which the g2SC gap $\De_3$ vanishes.
}
\label{fig:gaps_zero40}
\end{figure}

\begin{figure}
\includegraphics[width=8cm]{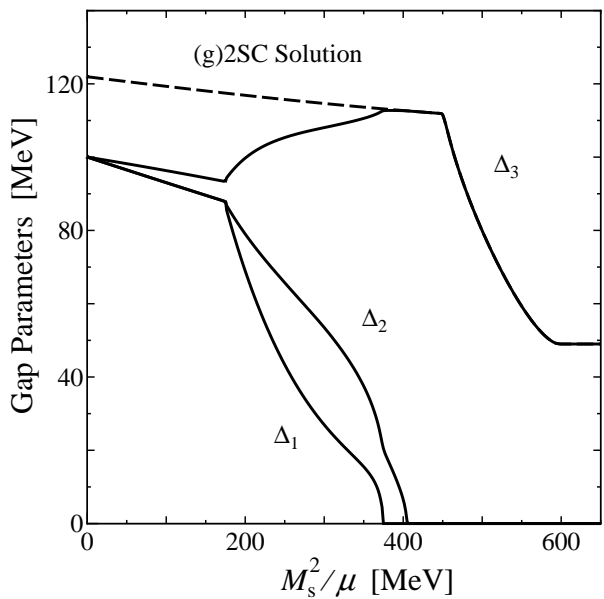}
\caption{Gap parameters for a still stronger interaction, with 
$\Delta_0=100\MeV$ for $\ms=0$. As in Fig.~\ref{fig:gaps_zero},
the CFL$\rightarrow$gCFL transition occurs where $\mssq\simeq 2\De_1$.
At $\mssq=375\MeV$, $\De_1$ vanishes at a second order 
gCFL$\rightarrow$uSC transition.
Then, at $\mssq=405\MeV$, $\De_2$ vanishes at a second order 
uSC$\rightarrow$2SC transition. At $\mssq=449\MeV$, there
is a second order
2SC$\rightarrow$g2SC transition. And finally,
for $\mssq>598\MeV$, corresponding to $M_s>547\MeV$, 
there are no more
strange quarks present in the system,
as shown explicitly in Fig.~\ref{fig:density}, 
and further increase in $M_s$
changes nothing.  
}
\label{fig:gaps_zero100}
\end{figure}

\begin{figure}
\includegraphics[width=8cm]{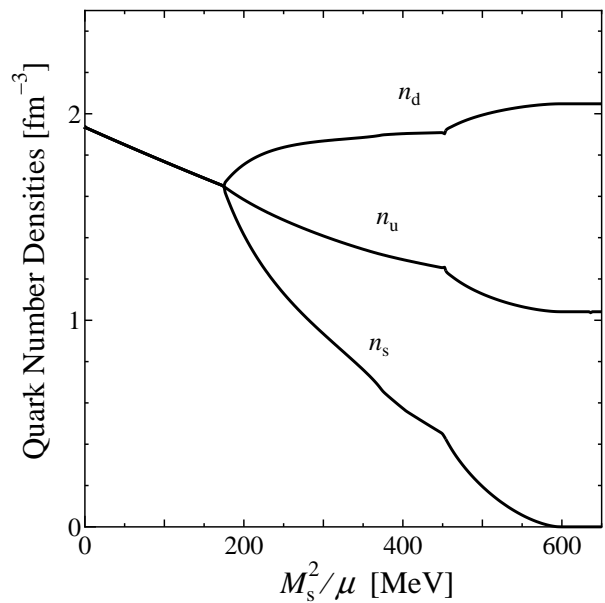}
\caption{Number densities of quarks with flavors $u$, $d$ and $s$
(in each case summed over the three colors) as a function
of $M_s^2/\mu$.  All parameters are as in Fig.~\ref{fig:gaps_zero100},
with $\De_0=100\MeV$.
We see that the number densities are equal only in the CFL phase,
and see that for $\mssq>598\MeV$ there are no strange quarks
present.
}
\label{fig:density}
\end{figure}

We now investigate how our zero temperature results
change if we vary the strength of the coupling, and hence $\De_0$.
In Fig.~\ref{fig:gaps_zero40}, we show the gap parameters
as a function of $\mssq$ with $\De_0=40\MeV$. We have changed
the scale on both the horizontal and vertical axes by the same
factor of $40/25$.  We see that the CFL$\rightarrow$gCFL transition
again occurs at $\mssq\simeq 2\De_1$, and that the shape of the 
curves in the gCFL region is qualitatively as before, when
suitably rescaled. However, at large values of $\mssq$
we now find a g2SC window: the gCFL free energy
crosses above that of the g2SC phase --- whose gap parameter
$\De_3$ is also shown in Fig.~\ref{fig:gaps_zero40} --- at the
vertical line in the figure, and the g2SC gap vanishes only at
a larger $\mssq$. (We have located the vertical line by comparing
free energies, as we did in Fig.~\ref{fig:energy_zero}, but we
shall not give the figure.)

If we reduce $\De_0$ from 25~MeV,
rather than increasing it, and rescale both axes of 
Fig.~\ref{fig:gaps_zero} by the same factor by which
we reduce $\De_0$, we obtain a figure that looks qualitatively
like Fig.~\ref{fig:gaps_zero}. We conclude that stronger
interaction tends to favor a g2SC window at large
values of $\mssq$, whereas weaker interaction disfavors it.
The boundary between the two cases is at $\De_0=25\MeV$ in our model.

It is interesting to ask what happens at still larger $\De_0$.
We show the gap parameters in our model with $\De_0=100\MeV$
in Fig.~\ref{fig:gaps_zero100}.  We see the CFL$\rightarrow$gCFL
transition at $\mssq=2\De_1$ once again.  The physics at 
and beyond the large-$\mssq$ boundary
of the gCFL regime is now qualitatively different.  This regime
corresponds either to very large values of $M_s$, or else to
such small values of $\mu$ that the hadronic phase will
likely have taken over, making the right half of this plot
somewhat academic.  One reason it is of interest, however,
is simply the fact that we can draw it: had
we made a small $M_s^2/\mu^2$ approximation 
as in Refs.~\cite{Alford:2003fq,Alford:2004hz,Ruster:2004eg},
this regime would be inaccessible. 
The figure shows 
a sequence of phases as $\mssq$ is increased
above the gCFL phase: (i) the gCFL phase ends at a second
order phase transition at which $\De_1\rightarrow 0$, above
which we find
a uSC window in which both $\De_2$ and $\De_3$ remain
nonzero; (ii) next, $\De_2\rightarrow 0$ at a second order
phase transition at which the uSC phase is succeeded by the
2SC phase; (iii) finally, there is a 2SC$\rightarrow$g2SC transition.
This sequence of phases agrees with
that found in Ref.~\cite{Ruster:2004eg} at large $\De_0$, in a 
calculation done using a small-$M_s^2/\mu^2$ approximation pushed
beyond its regime of validity.
Our present calculation
can be extended (within its model context) to arbitrarily large $M_s$.
Indeed, as can be seen in Fig.~\ref{fig:density}
we find a transition to two-flavor quark matter,
with zero strange quark density, at $M_s=547\MeV$, corresponding
to $\mssq=598\MeV$.  Below this $M_s$, we find the g2SC phase
with unpaired strange quarks. 
Above this value of $M_s$, 
we have two-flavor g2SC
quark matter and 
further increase in $M_s$ has no effect on the physics.

As $\De_0$ is increased from 40~MeV to 100~MeV, going from
Fig.~\ref{fig:gaps_zero40} to Fig.~\ref{fig:gaps_zero100},
the first qualitative change to occur is that the
gCFL phase ends at a second order
transition at which $\De_1\rightarrow 0$, 
instead of ending at a first order transition. 
Above this $\De_0$, the phase diagram includes a uSC window 
separated from the (g)2SC phase by a first order phase
transition.  At a somewhat larger $\De_0$, this first order
phase transition becomes second order. At a still larger $\De_0$,
the interaction is strong enough to have g2SC pairing in
the two flavor quark matter that our model describes
for $M_s\rightarrow\infty$, and the physics is as
in Fig.~\ref{fig:gaps_zero100}.


\section{The Phase Diagram at Nonzero Temperature}
\label{sec:nonzero}

\begin{figure}
\includegraphics[width=8cm]{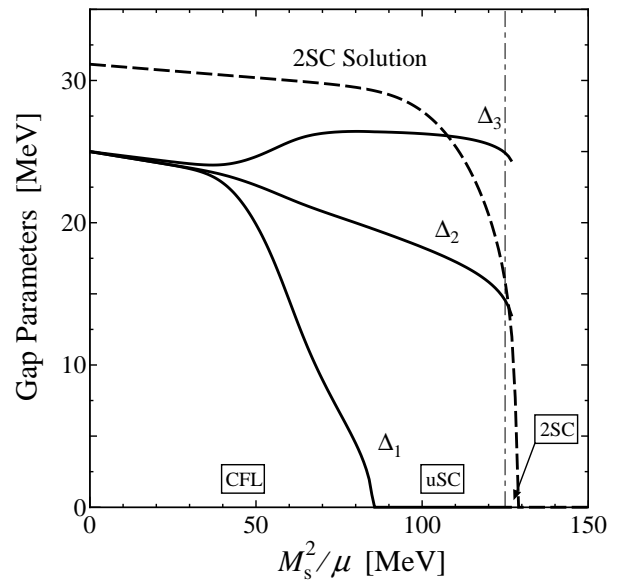}
\caption{Gap parameters as a function of $\mssq$
at $T=2\MeV$, with all other
parameters as in Fig.~\ref{fig:gaps_zero}.}
\label{fig:gaps_2}
\end{figure}

\begin{figure}
\includegraphics[width=8cm]{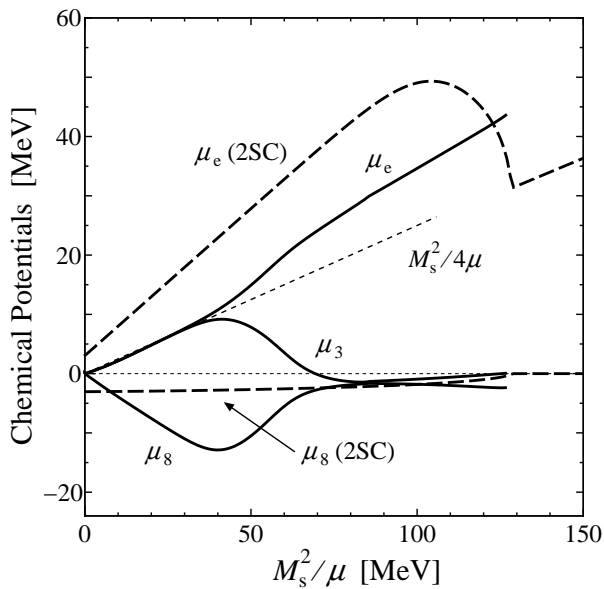}
\caption{Chemical potentials as a function
of $\mssq$ at $T=2\MeV$, with
all other parameters as in Fig.~\ref{fig:chem_zero}.
At small $\mssq$, $\mu_e$ and $\mu_3$ are close
to $M_s^2/4\mu$.}
\label{fig:chem_2}
\end{figure}

We now explore the solutions to the gap equations and neutrality
conditions at nonzero temperatures.  As in the previous
section, we begin with a coupling chosen so that $\De_0=25\MeV$.
The phase diagram for this value of the coupling is given
in Fig.~\ref{fig:phase_diagram}, to which the reader should refer
in this section.  We constructed Fig.~\ref{fig:phase_diagram}
by first making plots of the gap parameters and chemical potentials
versus $\mssq$ at many values of $T$, and versus $T$ at many
values of $\mssq$.  In this Section, we present and discuss
several of these
``sections'' of Fig.~\ref{fig:phase_diagram}, enough to understand
the many features of the phase diagram.  We then show phase diagrams
for $\De_0=40$ and $100\MeV$.

We start by turning on a small temperature,
$T=2\MeV$, and seeing how the plots of gaps and chemical potentials,
shown in Figs.~\ref{fig:gaps_2} and \ref{fig:chem_2}, change from
those at zero temperature, Figs.~\ref{fig:gaps_zero} 
and \ref{fig:chem_zero}.  We see many interesting changes
already at this relatively small temperature. The 
CFL$\rightarrow$gCFL transition seen at zero temperature in 
Figs.~\ref{fig:gaps_zero} and \ref{fig:chem_zero}
has become completely smooth at $T=2\MeV$:
there is no sharp difference between
CFL and gCFL at nonzero $T$. Furthermore, note that the 
zero temperature transition
from 2SC to g2SC is also washed out.  This makes sense: at $T=2\MeV$,
it makes no physical difference whether a certain fermionic quasiparticle
is gapless or has a gap that is nonzero but
smaller than 2~MeV. So, although in
Fig.~\ref{fig:phase_diagram} we have shown the values of $\mssq$
where quark quasiparticles become gapless within the CFL,
uSC and 2SC phases, these dashed lines have physical significance
only where they intersect $T=0$. 

We see in Fig.~\ref{fig:gaps_2} that $\De_1$ vanishes
at a second order CFL$\rightarrow$uSC transition at $\mssq=85.1\MeV$.
Another way to say this is that for $\mssq<85.1\MeV$, the critical
temperature at which $\De_1$ vanishes upon heating must be greater
than 2~MeV, whereas for $\mssq>85.1\MeV$, this critical temperature
is less than 2~MeV.  As at $T=0$, there is a first order phase
transition, denoted in Fig.~\ref{fig:gaps_2} 
by a thin vertical line, but here it is a first order phase transition
between the uSC phase at $\mssq<125\MeV$ and the 2SC phase at 
$\mssq>125\MeV$.  The 2SC phase ends at a second order
phase transition where $\De_3\rightarrow0$ at $\mssq=129\MeV$.
This means that there is a regime of $\mssq$ at which there
is 2SC pairing at $T=2\MeV$, but no
pairing at $T=0$. We investigate this further below.

\begin{figure}
\includegraphics[width=8cm]{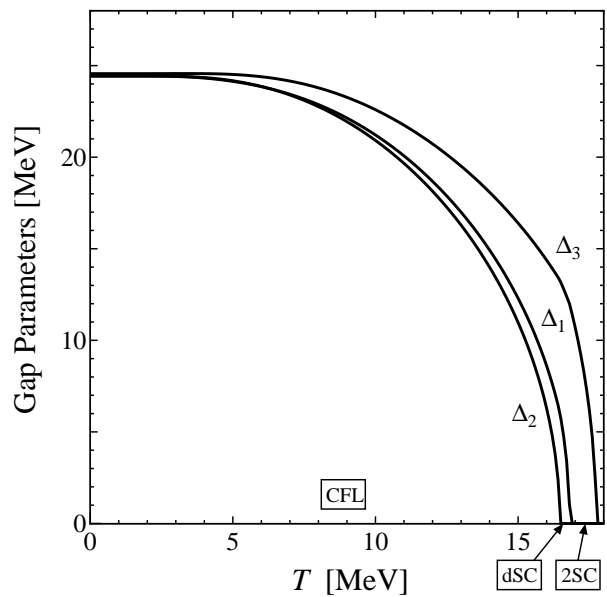}
\caption{Gap parameters as a function of $T$ at $\mssq=15\MeV$.}
\label{fig:gaps_dsc}
\end{figure}

\begin{figure}
\includegraphics[width=8cm]{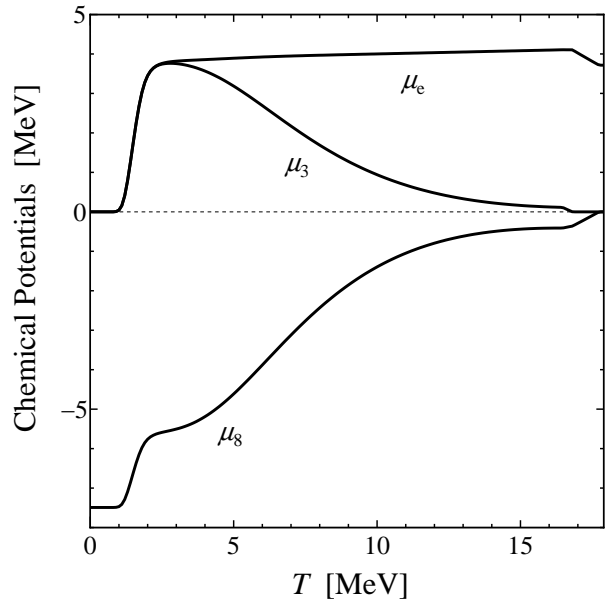}
\caption{Chemical potentials as a function of $T$ at $\mssq=15\MeV$.}
\label{fig:chem_dsc}
\end{figure}

We see in Fig.~\ref{fig:chem_2} that at small $\mssq$,
the chemical potentials
$\mu_e$ and $\mu_3$ are both close to $M_s^2/4\mu$ at $T=2\MeV$.
This is qualitatively
different than
their zero temperature behavior $\mu_e=\mu_3=0$.  In comparison,
$\mu_e$ takes on the value $M_s^2/4\mu$ in unpaired quark matter.
Thus, already at a temperature of only 2~MeV there is no sense in
which $\mu_e$ is small.  We see that at larger $\mssq$, $\mu_e$
and $\mu_3$ diverge as at zero temperature, but they do so 
smoothly and they diverge from $M_s^2/4\mu$, rather than
from 0.  Since we have found that the CFL phase has become a metal
already at $T=2\MeV$, it is natural to ask at what temperature
the insulator-metal crossover occurs. We answer this question
at $\mssq=15\MeV$ in Figs.~\ref{fig:gaps_dsc} and \ref{fig:chem_dsc}.
The latter figure shows a rapid insulator to metal crossover
occurring between $T=1\MeV$ and $T=2\MeV$, with $\mu_e$,
$\mu_3$ and $\mu_8$ all changing. We shall discuss this
crossover at length in Section~\ref{sec:crossover}. It can be understood
analytically, and we shall see in Section~\ref{sec:crossover}
that the reason that $\mu_e$
takes on the value $M_s^2/4\mu$ is quite different from
that in unpaired quark matter.  Above the crossover,
$\mu_e$ changes  little as $T$ increases
further but both $\mu_3$ and $\mu_8$
decrease in magnitude.  This occurs because at larger
temperatures the gap parameters decrease, as seen
in Fig.~\ref{fig:gaps_dsc}, and as the gap parameters
vanish color neutrality occurs with $\mu_3=\mu_8=0$~\cite{Iida:2003cc}
whereas electrical neutrality still requires a nonzero $\mu_e$.

We see in Fig.~\ref{fig:gaps_dsc} 
that the gap parameters change little at the low temperatures
at which the CFL phase is undergoing its insulator to metal
crossover.  Although it is not really visible in the figure,
we find that $\De_2>\De_1$ for $0<T<6.28\MeV$, and $\De_1>\De_2$
at higher temperatures.  At $T=16.46\MeV$, $\De_2$ vanishes at
a second order phase transition and we find the dSC phase.
Then, at $T=16.81\MeV$, $\De_1$ vanishes, yielding the 2SC phase.
The final phase transition, at which $\De_3$ vanishes,
occurs at $T=17.73\MeV$.  This ordering of phase transitions
is in qualitative agreement with that found in 
Ref.~\cite{Iida:2003cc}
using a Ginzburg-Landau approximation, and is in disagreement
with the results of Ref.~\cite{Ruster:2004eg}, in which no dSC
regime was found.  In order to gain
confidence in the accuracy of our calculation and in 
the existence of the dSC
phase, in Section~\ref{sec:GL} we make a detailed
and quantitative comparison between our results and those obtained
via the Ginzburg-Landau approximation.

\begin{figure}
\includegraphics[width=8cm]{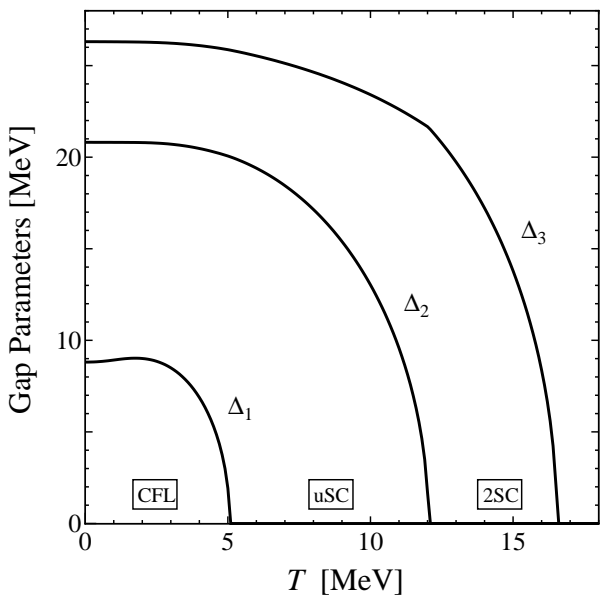}
\caption{Gap parameters versus temperature at $\mssq=70\MeV$.}
\label{fig:temp_usc}
\end{figure}

\begin{figure}
\includegraphics[width=8cm]{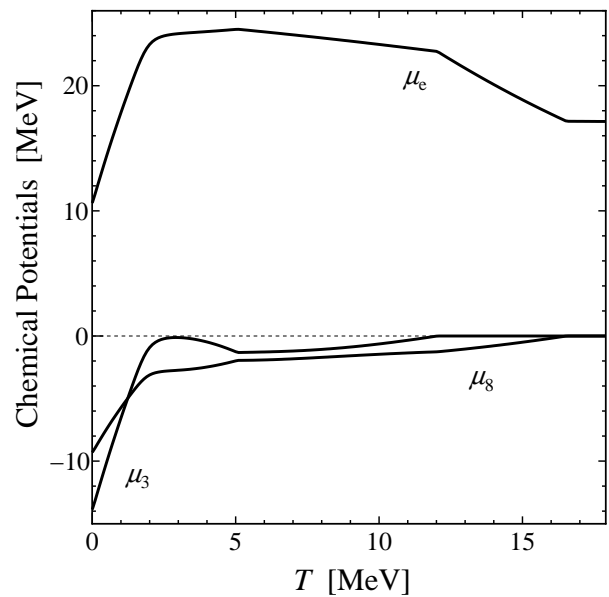}
\caption{Chemical potentials versus temperature at $\mssq=70\MeV$.}
\label{fig:chem_usc}
\end{figure}

At larger values of $\mssq$, the ordering of phase transitions
as a function of increasing temperature changes.  For example,
if we consider $\mssq=70\MeV$, in the gCFL phase at $T=0$ 
with $\De_1<\De_2<\De_3$, we see in Fig.~\ref{fig:temp_usc}
that as the temperature is increased, $\De_1$ vanishes first,
then $\De_2$ and then $\De_3$, meaning that the
phase which intervenes between CFL and 2SC is uSC, 
not dSC. This order of phase transitions is unsurprising,
given that $\De_1<\De_2<\De_3$ at $T=0$ in the gCFL phase.
All three transitions are
second order transitions in mean field theory. 
Fig.~\ref{fig:chem_usc} shows the chemical potentials
as a function of increasing temperature at $\mssq=70\MeV$.  
Whereas the $T=0$ CFL phase undergoes an insulator to
metal crossover as it is heated, the gCFL phase is already
a metal at $T=0$.

In both Figs.~\ref{fig:gaps_dsc} and \ref{fig:temp_usc},
we see a sequence of three second order phase transitions.
The first of these, a transition from the CFL phase
to either the dSC or the uSC phase, is likely not significantly
affected by gauge field fluctuations, because the
same gauge symmetries are unbroken (the $U(1)_{\Qtilde}$ symmetry)
and broken (the other eight gauge symmetries) on both sides
of the transition.  It is therefore an interesting question
for future work to consider the order parameter fluctuations
at this transition, asking whether they render it first order
or, if not, determining its universality class.  
The two mean field transitions occurring at higher temperatures in 
Figs.~\ref{fig:gaps_dsc} and \ref{fig:temp_usc}
will be qualitatively affected by gauge field fluctuations,
as at each of them there are gauge symmetries that are
broken on the low temperature side of the transition
and restored above the transition. Gauge field
fluctuations will presumably make these transitions first order,
and shift their critical temperatures upward.  These effects
will be significant, because the relevant gauge fields
are strongly coupled~\cite{Matsuura:2003md,Giannakis:2004xt}.


We have found that at small $\mssq$,
$\De_2$ vanishes at a lower temperature
than $\De_1$ whereas at larger $\mssq$, these two transitions
occur in the opposite order. There must, therefore, be some
$\mssq$ at which both vanish at the same temperature.
We see in Fig.~\ref{fig:phase_diagram} that this ``doubly
critical'' point occurs at $T=15.3\MeV$ and $\mssq=29.4\MeV$.

\begin{figure}
\includegraphics[width=8cm]{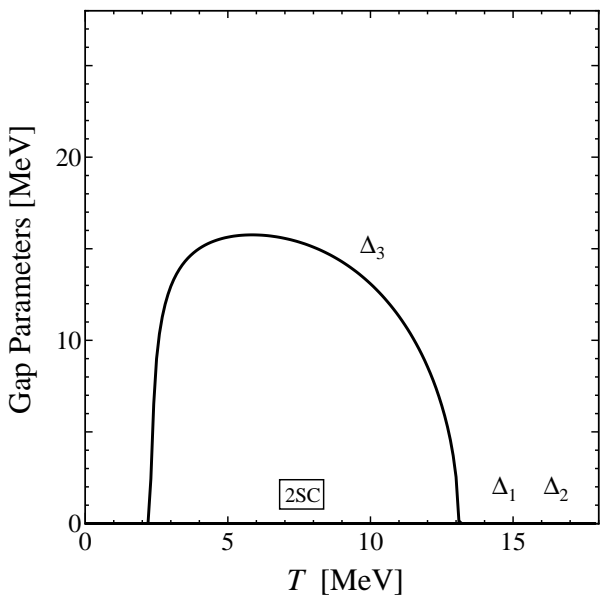}
\caption{Gap parameters versus temperature at $\mssq=130\MeV$.
At this large value of $\mssq$, both $\De_1$ and $\De_2$ are
zero at all temperatures. $\De_3=0$ at $T=0$, but this gap parameter
is nonzero for $2.37\MeV<T<13.08\MeV$.}
\label{fig:temp_2sc}
\end{figure}

Having followed what happens upon heating the CFL phase
at $\mssq=15\MeV$ and upon heating the gCFL phase at $\mssq=70\MeV$,
in Fig.~\ref{fig:temp_2sc}
we consider heating quark matter at $\mssq=130\MeV$,
which is unpaired at $T=0$. We see that $\De_3$ becomes
nonzero at a second order phase transition, and then
vanishes at a higher temperature at a second second order
phase transition.  This behavior has been described
previously~\cite{Sedrakian:1999cu,Shovkovy:2003uu}, 
and can be understood as follows.  At $T=0$,
the $u$ and $d$ Fermi surfaces in the unpaired quark matter
are too far apart to allow 2SC pairing.  However, as  
we increase $T$, we excite $u$ quarks above the $u$ Fermi surface,
and $d$ holes below the $d$ Fermi surface.  This smearing
of the separated Fermi surfaces
assists $u$-$d$
pairing, and $\De_3$ turns on at a nonzero temperature.  
Of course, at a still higher temperature the $\De_3$
condensate melts.

\begin{figure}
\includegraphics[width=8cm]{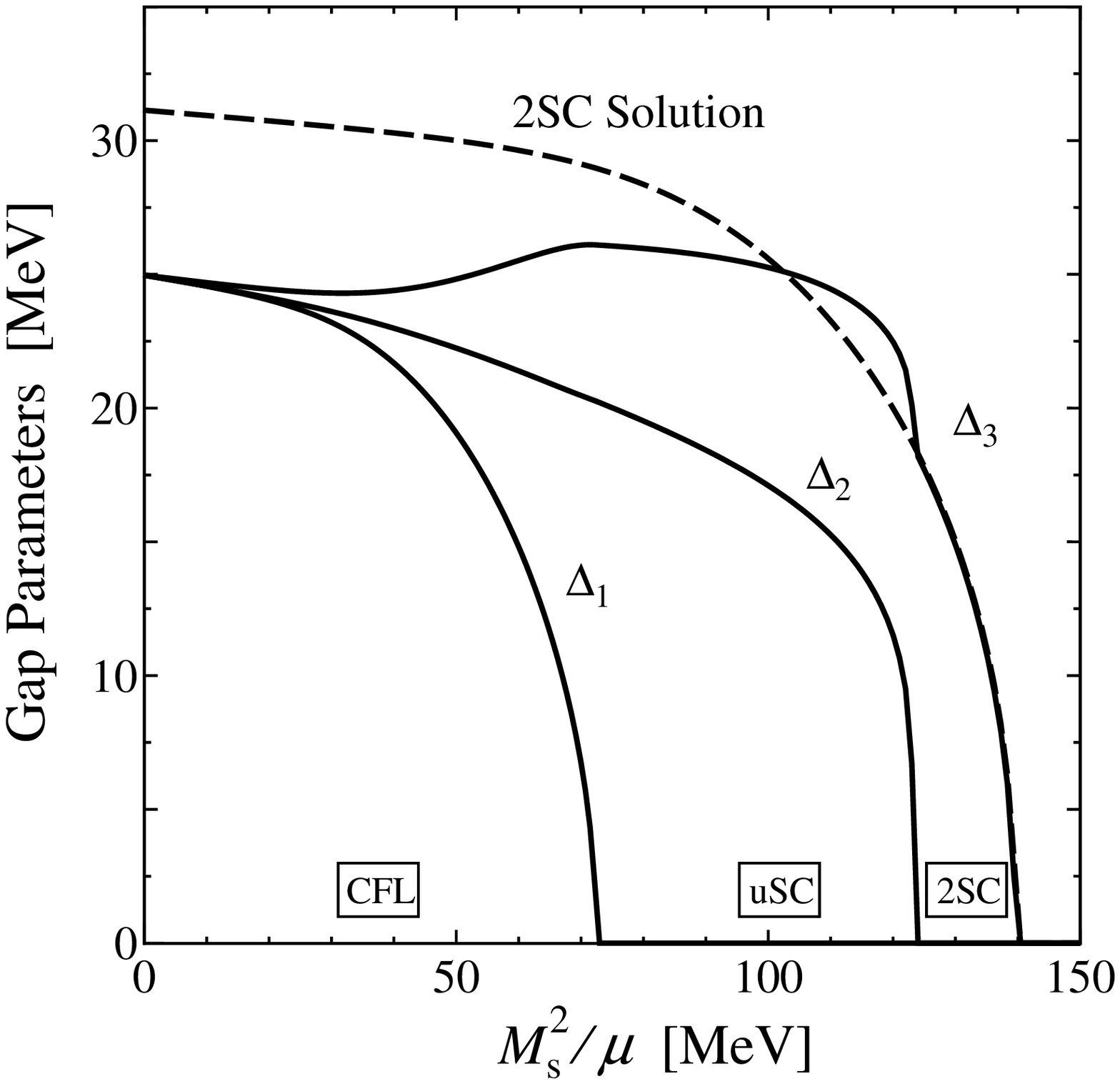}
\caption{Gap parameters versus $\mssq$ at $T=4\MeV$.}
\label{fig:gaps_4}
\end{figure}

The final slice of the phase diagram of Fig.~\ref{fig:phase_diagram}
that we shall show explicitly is a plot of the gap parameters
as a function of $\mssq$ at $T=4\MeV$, shown in Fig.~\ref{fig:gaps_4}.
By comparing this figure with Fig.~\ref{fig:gaps_2}, we see qualitative
changes in the physics between $T=2\MeV$ and $T=4\MeV$: the 2SC
phase now takes over from the uSC phase 
not via a first order phase transition, but instead
via a second order phase transition at $\mssq=123.2\MeV$ at 
which $\De_2\rightarrow 0$.  This means that a line of first
order phase transitions, present at lower temperatures, has
turned into a second order transition at a tricritical point.
This tricritical point is 
shown by a diamond in Fig.~\ref{fig:phase_diagram}, and
is located between $T=3.9\MeV$ and $T=4.0\MeV$.

\begin{figure*}
\includegraphics[width=12cm]{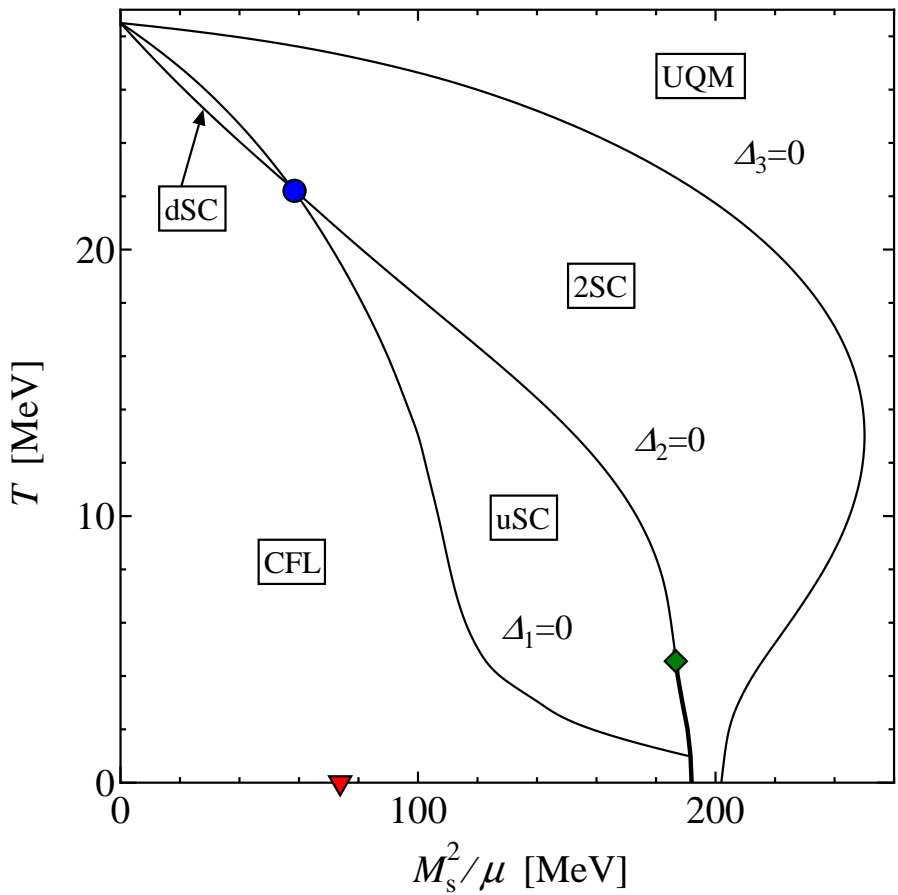}
\caption{Phase diagram of dense neutral quark matter in the 
$(M_s^2/\mu,T)$ plane, with a coupling chosen such
that $\De_0=40\MeV$. (All parameters except $\De_0$
are the same as in Fig.~\ref{fig:phase_diagram}.)
Only phase transitions are shown --- the analogues of the dashed
and dotted lines in Fig~\ref{fig:phase_diagram} are not given.
}
\label{fig:phase_diagram_40}
\end{figure*}

\begin{figure*}
\includegraphics[width=12cm]{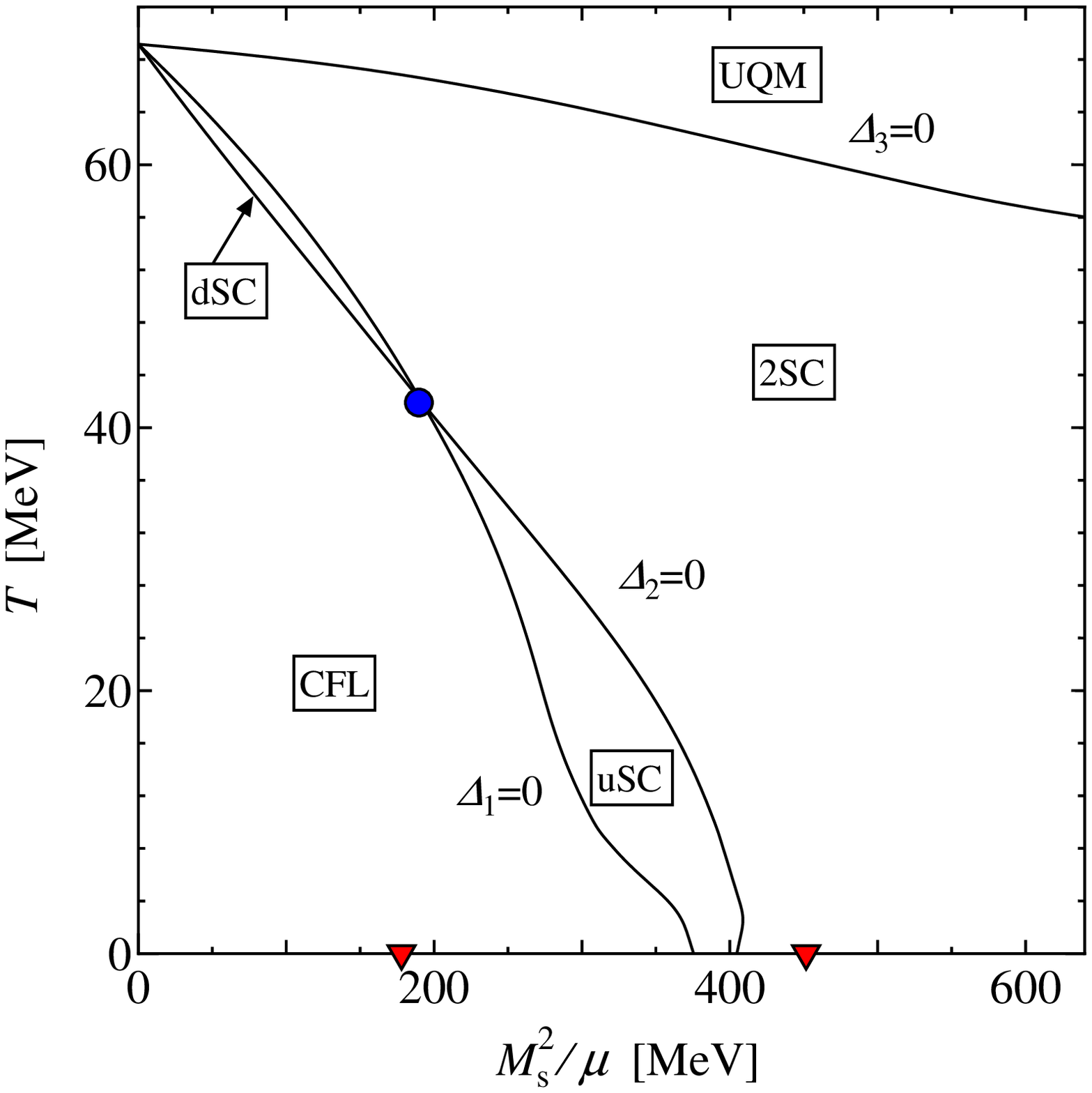}
\caption{Phase diagram of dense neutral quark matter in the 
$(M_s^2/\mu,T)$ plane, with a coupling chosen such
that $\De_0=100\MeV$. (All parameters except $\De_0$
are the same as in Figs.~\ref{fig:phase_diagram}
and \ref{fig:phase_diagram_40}.)  The two triangles
on the $T=0$ axis indicate quantum critical points at which 
modes become gapless. The first, which separates the
CFL and gCFL phases, is familiar. The second separates
2SC and g2SC phases, as shown in Fig.~\ref{fig:gaps_zero100}.
}
\label{fig:phase_diagram_100}
\end{figure*}

The results that we have presented up to
this point in this section,
plotted in Figs.~\ref{fig:gaps_2} 
through \ref{fig:gaps_4}, constitute a description of all
of the features at $T\neq0$ depicted in the phase diagram given
in Fig.~\ref{fig:phase_diagram}.  We now ask how this phase
diagram changes as we vary the strength of the interaction,
and hence $\De_0$.  If we reduce $\De_0$, there are no
qualitative changes as long as we rescale the vertical
and horizontal axes of the phase diagram by the same factor
that we reduce $\De_0$.  As at $T=0$, however,
increasing $\De_0$ leads to qualitative changes in the phase
diagram at large $\mssq$, indicating that the details of
the large $\mssq$ regions of the phase diagram are not robust
predictions of our model.   

With $\De_0=40\MeV$, in Fig.~\ref{fig:phase_diagram_40}, we see
the same three special points as in  Fig.~\ref{fig:phase_diagram}:
a quantum
critical point separating the CFL and gCFL phases 
at $T=0$, a doubly critical point at which the $\De_1\rightarrow 0$
and $\De_2\rightarrow 0$ transitions cross, and a tricritical
point at large $\mssq$ at which a line of first order phase
transitions ends.  Since both axes of Fig.~\ref{fig:phase_diagram_40}
have been rescaled by $40/25$ relative to Fig.~\ref{fig:phase_diagram},
the two figures are qualitatively similar:
the one qualitative change occurs at large $\mssq$, where the 
g2SC phase extends down to $T=0$, as we have already seen
in Fig.~\ref{fig:gaps_zero40}. (At $T=0$, there is a 
sharp distinction
between 2SC and g2SC, and in this instance the phase is g2SC.)
The most interesting quantitative change is a change
in the slopes of the $\De_1\rightarrow 0$ and  $\De_2\rightarrow 0$
transitions on the phase diagram at small $\mssq$, which pushes
the doubly critical point somewhat down in temperature.  
This effect is more clearly visible at stronger coupling,
with $\De_0=100\MeV$, as shown in Fig.~\ref{fig:phase_diagram_100}.
We shall  explain this quantitatively in Section~\ref{sec:GL}.  

With $\De_0=100\MeV$, in Fig.~\ref{fig:phase_diagram_100}, we
see further qualitative changes at large $\mssq$, corresponding
to those at $T=0$ shown in Fig.~\ref{fig:gaps_zero100}. Now,
the uSC phase and the 2SC phase both extend to $T=0$.  And, 
at $T=0$ there is a
2SC regime separated from g2SC by a quantum critical point, like
that separating the 
CFL and gCFL phases at $T=0$.
If we start at $T=0$ in the g2SC phase and heat the system,
$\De_3$ at first increases, before decreasing at higher
temperature, eventually vanishing at the upper phase transition
shown in Fig.~\ref{fig:phase_diagram_100}.  If we extended the
phase diagram to $\mssq\rightarrow\infty$, this critical temperature
would become $M_s$-independent in the
limit. The small remaining $M_s$-dependence
at the largest $\mssq$ we show is easily understood:
even though there are no strange quarks
present at $T=0$ at these large values of $M_s$, meaning that
the $T=0$ physics has become $M_s$-independent, strange quarks can
still be excited at nonzero temperature.

As $\De_0$ is increased from 40~MeV to 100~MeV, the phase
diagram changes continuously from that of Fig.~\ref{fig:phase_diagram_40}
to that of Fig.~\ref{fig:phase_diagram_100}.  First, the uSC
phase reaches the $T=0$ axis. Next, the tricritical point indicated
by the diamond in Fig.~\ref{fig:phase_diagram_40} retreats down
to $T=0$.  All the while, the (g)2SC region is extending farther
and farther to the right, eventually to $M_s\rightarrow\infty$
when the coupling is strong enough to allow $ud$ pairing
even once there are no strange quarks present.


We shall discuss the implications of the phase diagrams
that we have found in Section VII. First, however,
our investigation has raised several interesting questions that
we have been able to address analytically,
as we describe in Sections~\ref{sec:crossover} and \ref{sec:GL}.

\section{Heating the CFL phase: understanding the insulator 
to metal crossover}
\label{sec:crossover}

We have seen 
in Fig.~\ref{fig:chem_dsc} 
that as CFL quark matter is heated, it undergoes a crossover
from an insulator, with $\mu_e$ exponentially small, to
a metal, with $\mu_e\sim M_s^2/4\mu$.  With $\De_0=25\MeV$
and $\mssq=15\MeV$ as in 
Fig.~\ref{fig:chem_dsc},
the crossover occurs between 
$T=1\MeV$ and $T=2\MeV$.
This insulator to metal crossover has been seen previously
in Ref.~\cite{Ruster:2004eg}; 
our goal here is to understand it analytically.

We see from Fig.~\ref{fig:gaps_dsc}
that the gap parameters change little between $T=1\MeV$ and $T=2\MeV$,
whereas the chemical potentials change dramatically. We shall
explain the variation of the chemical potentials, treating the
gap parameters as $T$-independent.
We note from Fig.~\ref{fig:chem_dsc} that
during the crossover, $\mu_e$ increases from near 0 to
near $M_s^2/4\mu$, while $\mu_3$
increases by the same amount and
$\mu_8$ increases by half
as much. 
This tells us that (\ref{eq:CFL_mu3}) and (\ref{eq:CFL_mu8}) 
are maintained, and hence
it is the combination $\mu_\Qtilde$
of (\ref{muQtilde}) that is changing.  We recall that 
at $T=0$ the contribution of the quark matter to the
free energy $\Omega$ 
is independent of $\mu_\Qtilde$, and this ``plateau'' is only
curved by the small contribution of the electrons to the
free energy, of order $\mu_e^4$, which favors $\mu_e=0$.
Above the crossover we find
$\mu_e\simeq M_s^2/4\mu$, which is on the
plateau but away from the point on the plateau favored by
the electron neutrality condition.
In order to understand the crossover, then, there must be
thermally excited $\Qtilde$-charged quasi-particles
whose neutrality condition favors $\mu_e\neq 0$, and 
we must see the
curvature of the free energy plateau due to these
quasiparticles ``take over'' from that due to the electrons.
We first show that the CFL quark quasiparticle excitations
have
the desired effect, and then in Section~\ref{subsec:meson}
we show that the thermally excited charged mesonic
pseudo-Nambu-Goldstone boson excitations of the CFL phase
play a negligible role.

The quarks with nonzero $\Qtilde$ are the $rs$ and $bu$ which
pair with gap parameter $\De_2$, and the $rd$ and $gu$ which
pair with gap parameter $\De_3$. This means that in the CFL
phase there are two quasiquarks with $\Qtilde=+1$ (one
a linear combination of $bu$ quarks and $rs$ holes;
the other a linear combination of $gu$ quarks and $rd$ holes)
and two quasiquarks with $\Qtilde=-1$ (one 
a linear combination of $rs$ quarks and $bu$ holes;
the other a linear combination of $rd$ quarks and $gu$ holes).

We now evaluate the dispersion relations of these excitations,
and estimate the number density of thermally excited charged
quasiquarks. In this Section,
we shall follow Refs.~\cite{Alford:2003fq,Alford:2004hz} and include the nonzero
strange quark mass only via its effect as a shift in the
chemical potentials of the strange quarks. We have
seen in Section~\ref{sec:zero} that this is a good approximation in
the CFL phase. The dispersion relations of these four quasiparticles 
are given by~\cite{Alford:2004hz}
\begin{align}
 \varepsilon_{(rs\text{-}bu)}(p) &= 
  \pm 
\half\left(\mu_{rs}-\frac{M_s^2}{2\mu}-\mu_{bu}\right)\nonumber\\
&\qquad +
\sqrt{\left(p-\bar\mu_{(bu\text{-}rs)}
+\frac{M_s^2}{4\mu}\right)^2+\Delta_2^2}\nonumber\\
 \varepsilon_{(rd\text{-}gu)}(p) &= 
  \pm \half(\mu_{gu}-\mu_{rd})
+\sqrt{(p-\bar\mu_{(rd\text{-}gu)})^2+\Delta_3^2}
\end{align}
where 
$\bar\mu_{bu\text{-}rs}\equiv (\mu_{bu}+\mu_{rs})/2$ and
$\bar\mu_{rd\text{-}gu}\equiv (\mu_{rd}+\mu_{gu})/2$.
Upon substituting the definitions (\ref{mudefinitions}) and the relations
(\ref{eq:CFL_mu3}) and (\ref{eq:CFL_mu8}), which are maintained
through the crossover, these become 
\begin{align}
 \varepsilon_{(rs\text{-}bu)}(p) &= 
  \pm \left(\mu_e-\frac{M_s^2}{2\mu}\right)
+\sqrt{\left(p-\mu
+\frac{M_s^2}{6\mu}\right)^2+\Delta_2^2}\nonumber\\
 \varepsilon_{(rd\text{-}gu)}(p) &= 
  \pm \mu_e
+\sqrt{\left(p-\mu+\frac{M_s^2}{6\mu}\right)^2
+\Delta_3^2}\ .
\end{align}
At the temperatures of interest, these excitation energies 
are all greater than $T$, and so only the lowest energy excitation
with each $\Qtilde$ charge matters --- the 
number density of the higher energy excitations is
exponentially smaller. Labeling the quasiparticles by 
their $\Qtilde$ charge, the excitation energies of the
lowest lying charged quasiquarks are
\begin{align}
 \varepsilon_{+1}(p) &= 
  \left(\mu_e-\frac{M_s^2}{2\mu}\right)
+\sqrt{\left(p-\mu
+\frac{M_s^2}{6\mu}\right)^2+\Delta_2^2}\nonumber\\
 \varepsilon_{-1}(p) &= 
  - \mu_e
+\sqrt{\left(p-\mu+\frac{M_s^2}{6\mu}\right)^2
+\Delta_3^2}\ ,
\end{align}
corresponding to the ($rs$-$bu$) quasiquark with $\Qtilde=+1$ 
and the ($rd$-$gu$) quasiquark with $\Qtilde=-1$.
We now evaluate the $T$-dependent contribution of
these quasiquarks to the free energy $\Omega$
given by (\ref{eq:potential}).  Because the quasiquark energies are
much larger than the temperatures of interest, the
Boltzmann factors are small, the 
integral is dominated by $p$ near the minimum of $\varepsilon(p)$,
and we can use the saddle-point approximation.  
We find that, for example,
the contribution of the $\Qtilde=+1$ quasiquark to $\Omega$
is given by
\begin{align}
 & -\frac{1}{\pi^2}
\int_0^\Lambda dp\; p^2 T \ln\bigl(1+\e^{-|\varepsilon_{+1}(p)|/T}
  \bigr) \notag\\
 \simeq & -\frac{1}{\pi^2}
\int_0^\Lambda dp\; p^2 T\, e^{-|\varepsilon_{+1}(p)|/T}
   \notag\\
 \simeq & -\frac{1}{\pi^2}\sqrt{2\pi\Delta T}\bar\mu^2 T
   e^{-|\varepsilon_{+1}(\bar\mu)|/T} \ ,
\end{align}
where $\bar\mu\equiv \mu-M_s^2/6\mu$. The contribution of
the $\Qtilde=-1$ quasiparticle is analogous.
The contribution of these two quasiparticles to the
$\Qtilde$-charge density is then
\begin{align}
 (\Qtilde=+1) &&& \frac{\sqrt{2\pi\Delta T}\bar\mu^2}{\pi^2}
  e^{-(\Delta+\mue-\ms^2/2\mu)/T}, \\
 (\Qtilde=-1) &&& -\frac{\sqrt{2\pi\Delta T}\bar\mu^2}{\pi^2}
  e^{-(\Delta-\mue)/T} .
\end{align}
We
have set $\De_3=\De_2\equiv \De$, a
good approximation throughout the crossover 
as shown in Fig.~\ref{fig:gaps_dsc}.
$\Qtilde$-neutrality is a balance between the charge densities
of these two quark quasiparticles and the electrons.

We can now see that what drives the system to $\mu_e\neq 0$
is the fact that the lightest $\Qtilde=+1$ and $\Qtilde=-1$ quark
quasiparticles have dispersion relations with different 
gaps when $M_s\neq 0$, and
consequently contribute charge densities of different magnitudes
when thermally excited.  If we
attempt to
set $\mu_e=0$ at $T\neq0$, there are more $\Qtilde=+1$ quasiparticles
present than $\Qtilde=-1$ quasiparticles, and the system is not neutral.
To achieve neutrality, $\mu_e$ must be increased as this increases 
the density of $\Qtilde=-1$ quasiparticles, decreases
the density of $\Qtilde=+1$ quasiparticles, and adds
electrons, which have $\Qtilde=-1$. This is described
by the neutrality condition 
\begin{align}
 & \frac{2\sqrt{2\pi\Delta T}\bar\mu^2}{\pi^2}\e^{-(\Delta-\ms^2/4\mu)/T}
  \sinh\biggl(\frac{\ms^2/4\mu-\mue}{T}\biggr) \notag\\
 & \qquad\qquad\qquad\qquad\qquad
  -\frac{\mue^3}{3\pi^2}-\frac{\mue T^2}{3}=0 \ ,
\label{eq:analytic}
\end{align}
which we can solve for $\mu_e(T)$ if we take $\De$ to be $T$-independent.

\begin{figure}
\includegraphics[width=8cm]{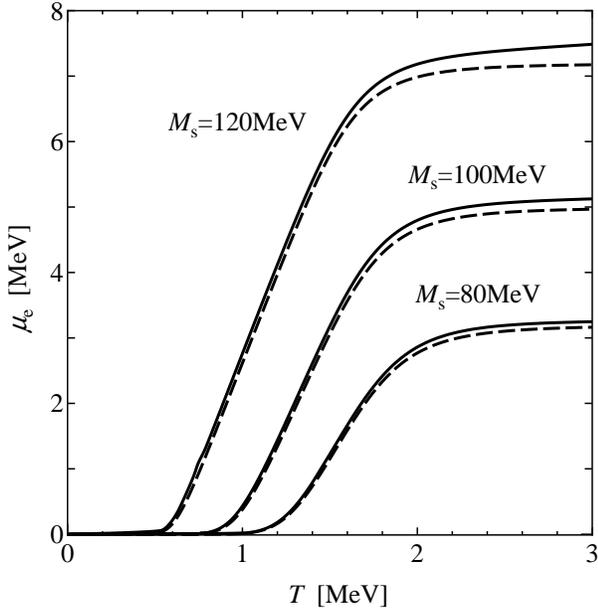}
\caption{Comparison between the analytic results (dashed curves) and
the numerical results (solid curves) for $\mu_e$ as a function
of temperature with $\De_0=25\MeV$ for several values of $M_s$ in the
CFL phase.  The analytic results were obtained from
(\ref{eq:analytic}), and the numerical results were obtained
by solving the full coupled gap equations and neutrality
conditions, as in previous Sections.  In evaluating $\mu_e$ from
(\ref{eq:analytic}) we have taken $\De$ to be the average of
$\De_2$ and $\De_3$ at the midpoint of the crossover.
}
\label{fig:comparison}
\end{figure}

Let us now investigate the implications of this result.
At very small $T$, the quark quasiparticles are exponentially
rare, and those with $\Qtilde=-1$ are exponentially rarer
than those with $\Qtilde=+1$.  The $\Qtilde=+1$ quasiparticle density
is balanced by the electron density, and $\mu_e$ is exponentially
small.  However, the quasiparticle densities are proportional
to $\mu^2$ whereas the electron density is not.  This means that
at the $T$ at which $\mu_e$ starts to take off, the 
quasiparticle Boltzmann factors are still rather small.  Once $T$
is large enough that $\mu_e$
approaches $M_s^2/4\mu$, however, 
even though the individual Boltzmann factors continue to rise
rapidly as $T$ increases further, 
the $\sinh$ factor in (\ref{eq:analytic})
becomes small.  Neutrality at this point
is primarily a balance between the densities of the $\Qtilde=+1$
and $\Qtilde=-1$ quasiparticles, with electrons cancelling
only the small difference between their densities.  The result,
seen already in Fig.~\ref{fig:chem_dsc} and shown in greater detail
in Fig.~\ref{fig:comparison} is a crossover in which $\mu_e$
is at first exponentially small, then rises rapidly, and then
saturates as it approaches $M_s^2/4\mu$. 
We see in 
Fig.~\ref{fig:comparison} that the equation (\ref{eq:analytic}
for $\mu_e$ that we have derived , making approximations as described, 
gives a very
good description of the numerical solution of the full coupled gap
and neutrality equations.  This demonstrates that
(\ref{eq:analytic}) provides us with a good analytic description
of the insulator to metal crossover that CFL quark matter
experiences when heated.

We have set the electron
mass to zero in Fig.~\ref{fig:comparison} and throughout. Including
it means that it takes a larger $\mu_e$ to achieve a given electron
density, pushing all the curves in Fig.~\ref{fig:comparison}
very slightly upwards.
With an electron mass as in nature, the effect on the curves
is invisible on the scale of the plot.

\subsection{Contribution of charged mesons}
\label{subsec:meson}
 
\begin{figure}
\includegraphics[width=8cm]{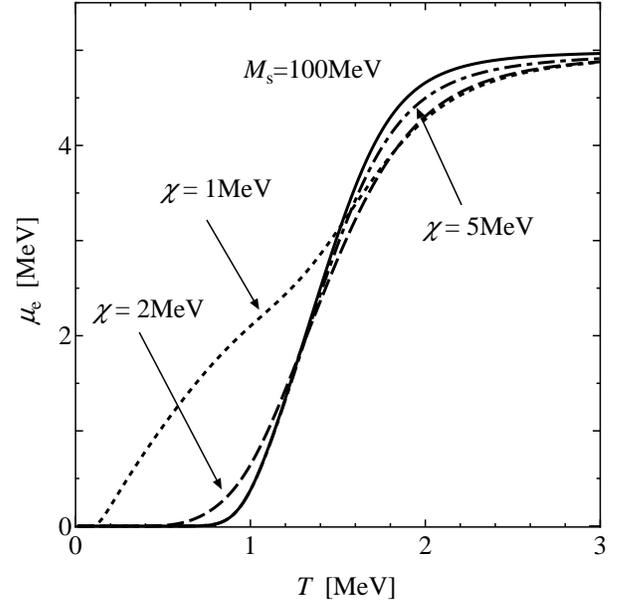}
\caption{Comparison between the analytic estimate
for $\mu_e$ versus $T$ with and without
charged meson contributions.  The solid curve, with
no meson contributions, is the same as the $M_s=100\MeV$
analytic estimate in Fig.~\ref{fig:comparison}.
The dashed  curves include the contribution of thermally
excited
charged pions and kaons, for three values  of the parameter
$\chi$ described in the text that parameterizes the
instanton contribution to the meson masses.
}
\label{fig:comparison_meson}
\end{figure}

As described in Section~\ref{subsec:symmetries}, there are other charged excitations
in the CFL phase. Among the pseudo-Nambu-Goldstone bosons, there
are mesons with the quantum numbers of the $\pi^\pm$ and $K^\pm$.
We have neglected the contribution of thermally
excited charged mesons to the charge density in 
the derivation of (\ref{eq:analytic}). We now 
investigate this approximation.

The dispersion relations of the charged mesons, together
with that for the $K^0$-meson which we shall
also need below, are given 
by~\cite{Son:1999cm,Hong:1999ei,Manuel:2000wm,Schafer:2002ty,Bedaque:2001je,Reddy:2002xc}
\begin{align}
\varepsilon_{\pi^{\pm}}(p) &= \pm \mu_e 
+\sqrt{v^2 p^2 + M^2_{\pi^{\pm}}}\nonumber\\
\varepsilon_{K^{\pm}}(p) &= \pm \mu_e 
\mp \frac{M_s^2}{2\mu}+\sqrt{v^2 p^2 + M^2_{K^{\pm}}}
\nonumber\\
\varepsilon_{K^{0}}(p) &=  - \frac{M_s^2}{2\mu}+\sqrt{v^2 p^2 + M^2_{K^{0}}}\ ,
\label{eq:mesondispersions}
\end{align}
where $v^2=1/3$ at high density~\cite{Son:1999cm}.
The meson masses in the CFL phase are given by
\begin{align}
M^2_{\pi^\pm} &= a(M_u+M_d)M_s + \chi(M_u+M_d)\nonumber\\
M^2_{K^\pm} &= a(M_u+M_s)M_d + \chi(M_u+M_s)\nonumber\\
M^2_{K^0} &= a(M_d+M_s)M_u + \chi(M_d+M_s) \ .
\label{eq:mesonmasses}
\end{align}
Here, $a=3\De^2/\pi^2 f_\pi^2$ with 
$f^2_\pi=(21-8\log 2)\mu^2/36\pi^2$ 
at high density~\cite{Son:1999cm}, which yields
$a=0.0175$ for $\De_0=25\MeV$ at $\mu=500\MeV$.
And, $\chi$ parameterizes
the contribution of $U(1)_A$-breaking
instanton effects which generate $\langle \bar q q \rangle$ condensates
and therefore contributions to meson masses
in the CFL phase~\cite{Manuel:2000wm,Schafer:2002ty,Reddy:2002xc}, 
via the 't~Hooft interaction which  
contributes couplings of the form $\De^2\langle \bar q q \rangle$.
The magnitude of $\chi$ is not well known, as it depends on the
instanton size distribution and instanton form factors at nonzero density.
It has been estimated to lie in the 
range $1\MeV<\chi<100\MeV$~\cite{Schafer:2002ty,Reddy:2002xc}.
In Fig.~\ref{fig:comparison_meson}, we include the contribution
of all four charged mesons to the $\Qtilde$-charge neutrality
condition, determining the density of thermally excited
bosonic quasiparticles from the dispersion relations. 
The contribution to the charge density from the $K^\pm$
mesons is
\begin{equation}
 \frac{1}{2\pi^2}\int_0^\infty \!\!\! dp\; p^2\biggl(
  \frac{1}{\exp\left[\varepsilon_{K^+}(p)/T\right]\!-\!1}
  -\frac{1}{\exp\left[\varepsilon_{K^-}(p)/T\right]\!-\!1}\biggr),
\end{equation}
and that from the pions is analogous.
We then solve for $\mu_e$ vs. $T$,
taking $M_s=100\MeV$ and $M_u=5\MeV$ and $M_d=10\MeV$. 
We plot the results for $\chi=$ 1, 2 and 5 MeV 
in Fig.~\ref{fig:comparison_meson}.

Adding the charged mesons adds new charge carriers, and so
the simplest expectation for their effects is that a smaller $\mu_e$
will be required in order to neutralize the imbalance in the fermionic
quasiparticle sector.  We see in Fig.~\ref{fig:comparison_meson}
that this expectation is borne out above the crossover, but not
below.  Above the crossover, we see that $\mu_e$ is reduced relative
to the results we obtained  in Fig.~\ref{fig:comparison},
where we neglected the mesons.  The mesonic contributions
get less significant at larger values of $\chi$, 
as the mesons get heavier. They are already
small for $\chi=1\MeV$ and are negligible by $\chi=5\MeV$.
At low temperatures, and for the smallest values of $\chi$,
there is another effect to be understood. The dispersion relations for 
the $K^\pm$ in (\ref{eq:mesondispersions}) indicate that
the $K^+$ is easier to excite than the $K^-$. This means
that they behave like the quasiquarks, in the sense that at 
a nonzero temperature they contribute a positive $\Qtilde$-charge
density, which must be cancelled.  If $\chi$ is very small,
the $K^+$ charge density becomes significant at such a low temperature
that its contribution can only be cancelled by electrons --- the
Boltzmann factors for all other excitations are still prohibitive.
This means that if $\chi$ is very small, $\mu_e$ initially rises 
with temperature significantly more
rapidly than in the absence of the mesons.  
We see this effect clearly in Fig.~\ref{fig:comparison_meson} for 
$\chi=1\MeV$ and still to a small degree for $\chi=2\MeV$.
For $\chi=5\MeV$, $\mu_e(T)$ is indistinguishable at low temperatures
from that in the absence of the mesons.  

We can summarize the results in Fig.~\ref{fig:comparison_meson} as follows.
For $\chi\gtrsim 2 \MeV$, the thermally excited charged mesons
have no significant effects at any temperature. But, if
$\chi\sim 1 \MeV$, which is at the bottom end of the estimated allowed
range $1\MeV < \chi <100\MeV$~\cite{Schafer:2002ty,Reddy:2002xc},
the $K^+$ excitations contribute significantly
at very low temperatures.

At much larger temperatures, well above the insulator to metal
crossover that we have analyzed, we come to the various
critical temperatures that we have seen in Section~\ref{sec:nonzero} and
will analyze further in Section~\ref{sec:GL}. At these temperatures,
at which gap parameters vanish, it is well understood that
the fermionic quasiparticles are the most important degrees
of freedom. What we have shown is that at low temperatures also, the
charged mesons are less important than the charged fermions among
the thermal excitations, as long as $\chi$ is not very small.

Mesons can nevertheless play an important
role if they condense~\cite{Bedaque:2001je}.  
At $T=0$, the CFL phase
is stable against meson condensation as long
as $\varepsilon_{K^0}>0$ at $p=0$.  
($K^0$-condensation
yields the most stringent constraint.)  This requires
$M_{K^0}^2>M_s^4/4\mu^2$, corresponding to $\chi\gtrsim M_s^3/4\mu^2$.
For $M_s=100\MeV$, as in Fig.~\ref{fig:comparison_meson}, this
requires $\chi>1\MeV$.  We are therefore justified in
our neglect of $K^0$-condensation in 
Fig.~\ref{fig:comparison_meson}.  If $\chi>3.6\MeV$,
there is no $K^0$-condensation in the CFL phase
with $\De_0=25\MeV$ for $M_s^2/\mu<46.8\MeV$,
meaning that there is no $K^0$-condensation
at $T=0$ for all values of $\mssq$ below the CFL$\rightarrow$gCFL
transition.  ($K^0$-condensation in the gCFL phase has
yet to be analyzed.)
It seems likely that $\chi>3.6\MeV$ at accessible densities,
since $\chi$ is larger at lower densities and the 
analysis in Ref.~\cite{Schafer:2002ty} which yields the
estimate $1\MeV<\chi<100\MeV$ becomes more reliable
at higher densities.  If the coupling is stronger, however,
$K^0$-condensation becomes more likely. For example,
for $\De_0=100\MeV$, there is no $K^0$-condensation
in the CFL phase only if $\chi>26\MeV$.  
If $K^0$-condensation were to
occur, it delays the CFL$\rightarrow$gCFL transition, increasing
the $M_s^2$ at which it occurs by a factor 4/3 
if $\chi=0$~\cite{Kryjevski:2004jw}. Further work remains to be done,
for example extending the analysis of 
Ref.~\cite{Kryjevski:2004jw} to nonzero $\chi$
and to the gCFL phase.  

Another issue that remains to be investigated
is the possibility of $\pi^-$ condensation at nonzero temperature.
We have seen that 
above the insulator to metal transition, $\mu_e\approx M_s^2/4\mu$.
At zero temperature, this would lead to $\pi^-$ condensation
if $\chi < M_s^4/(16 \mu^2 (M_u+M_d))$, but this $\mu_e$
only arises at $T\neq 0$, and nonzero $T$ acts to stabilize
against meson condensation.

Finally, all these issues should be investigated in an expanded
model in which the 't~Hooft interaction is included from the
beginning in the free energy and hence in
the gap and neutrality conditions, and the quark
masses are also solved for dynamically.

\section{The Ginzburg-Landau approximation}
\label{sec:GL}

In the $\mssq \rightarrow 0$ limit, the three critical temperatures
at which the gap parameters vanish become one. Near this critical
temperature, where all gap parameters are small, a Ginzburg-Landau
approximation can be employed.  The Ginzburg-Landau free energy
was analyzed at $M_s=0$ in Refs.~\cite{Iida:2000ha,Iida:2002ev}, 
and was extended
to small but nonzero $\mssq$ in Ref.~\cite{Iida:2003cc}.  We wish
to compare our results at small but nonzero $\mssq$
to those obtained in the Ginzburg-Landau
approximation, and to compare the coefficients 
in the Ginzburg-Landau potential (actually,
ratios of coefficients) in our model to those calculated in QCD
at asymptotic densities and thus at
weak coupling in Refs.~\cite{Iida:2000ha,Iida:2003cc}.  

The Ginzburg-Landau potential is parameterized as~\cite{Iida:2003cc}
\begin{equation}
 \begin{split}
 \Omega =& \alpha\bigl(\Delta_1^2+\Delta_2^2+\Delta_3^2\bigr)\\
  & +\epsilon\bigl(\Delta_1^2+\Delta_2^2\bigr) 
+ \third\eta \bigl(-2\Delta_1^2 +\De_2^2+\De_3^2\bigr) \\
 & +\beta_1\bigl(\Delta_1^2+\Delta_2^2+\Delta_3^2\bigr)^2
   +\beta_2\bigl(\Delta_1^4+\Delta_2^4+\Delta_3^4\bigr),
 \end{split}
\label{eq:GL}
\end{equation}
where 
$\alpha=\alpha_0(T-T_{\rm c})/T_{\rm c}$ and where the coefficients
$\epsilon$ and $\eta$ are proportional to $M_s^2$ and $\mu_e$ respectively
and are therefore present only if $M_s\neq 0$.  The form of
the $\epsilon$ and $\eta$ terms was derived in Ref.~\cite{Iida:2003cc}
for QCD at asymptotic densities. In the Ginzburg-Landau limit, in 
which all gap parameters are small, color neutrality occurs
with $\mu_3$ and $\mu_8$ vanishingly small~\cite{Iida:2003cc},
and electrical neutrality requires a nonzero $\mu_e$, of order
$M_s^2/4\mu$ as in unpaired quark matter.  At low temperatures,
the most important consequences of $M_s$ and $\mu_e$ (and also $\mu_3$ and 
$\mu_8$) derive from the stress they put on pairing, as  they
(seek to) push Fermi momenta apart.  Near $T_c$, $T\gg \De$. And,
we are working at 
$M_s^2/\mu\ll T_c$, where the thermal smearing of the Fermi distributions is
much greater than the splittings between Fermi momenta.  
In this regime, the effects of $M_s^2$ or $\mu_e$ 
on the pairing of two quarks
do not arise from the splitting between their Fermi momenta,
as this is negligible.  
The effects of $M_s^2$ or $\mu_e$
arise from the change in the {\it average} Fermi momenta
of the two quarks,
and hence in the density of states, that $M_s^2$ or $\mu_e$ induces. 
For example, 
$M_s^2$ depresses
the average Fermi momenta of $u$-$s$ and $d$-$s$ pairs, but does
not affect $u$-$d$ pairs.  This explains the form of the $\epsilon$ term in
(\ref{eq:GL}). On the other hand, $\mu_e>0$ increases the average
Fermi momenta of $d$-$s$ pairs, while decreasing that of
$u$-$d$ and $u$-$s$ pairs.  This explains the form of the $\eta$ term.

\begin{figure}
\includegraphics[width=4cm]{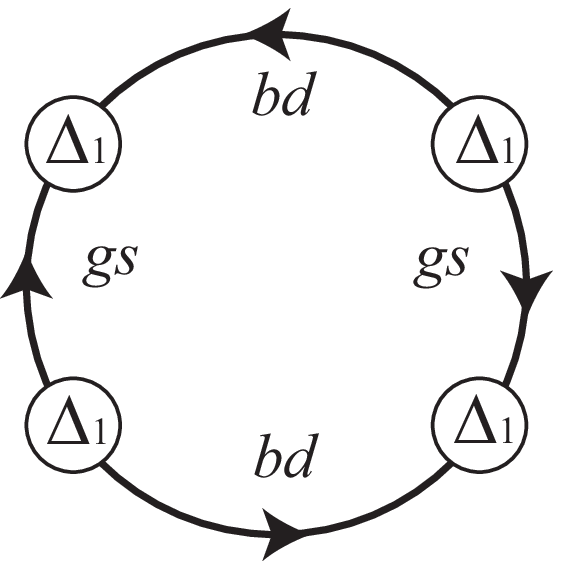}
\includegraphics[width=4cm]{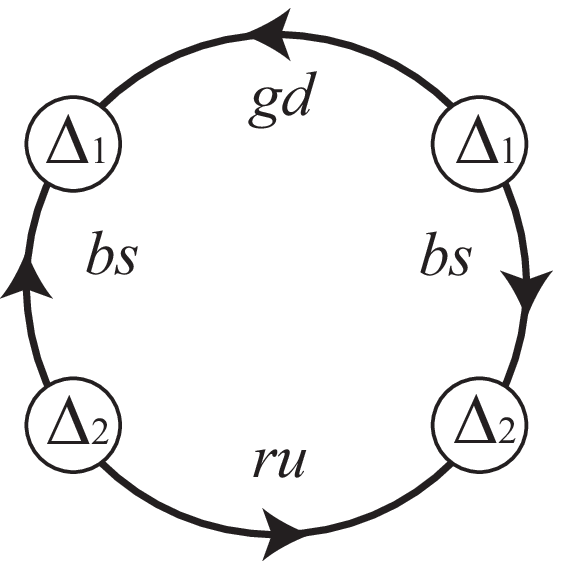}
\caption{Diagrams contributing to the $\De_1^4$ and $\De_1^2\De_2^2$
terms in the Ginzburg Landau potential (\ref{eq:GL}).}
\label{fig:diagrams4}
\end{figure}

In Ref.~\cite{Iida:2000ha} it is shown that, in the 
weak coupling regime, $\beta_1=\beta_2$.  This result is valid
beyond weak coupling, however, as all that is required to demonstrate
it is the pairing ansatz (\ref{blocks}) and the mean field approximation.
The quarks in the $2\times 2$ blocks contribute through diagrams
like the first in Fig.~\ref{fig:diagrams4}, leading to a contribution
proportional to $\De_1^4+\De_2^4+\De_3^4$.  The quarks in the $3\times 3$
block contribute through diagrams like both the first and second in
Fig.~\ref{fig:diagrams4}, leading to a contribution proportional to
$\De_1^4+\De_2^4+\De_3^4+2\De_1^2\De_2^2+2\De_2^2\De_3^2+2\De_1^2\De_3^2$.
Adding all the diagrams, we conclude that $\beta_1=\beta_2$.

We now wish to confirm that (\ref{eq:GL}) correctly describes
the $M_s$-dependent physics in our model.  We do so by extracting
the ratio 
\begin{equation}
\zeta\equiv\frac{\epsilon}{\eta}
\end{equation}
from our results in three independent ways. 
If (\ref{eq:GL}) is correct --- that is if there are no $M_s$-dependent
terms missed --- the three extractions of $\zeta$ should agree.  
We see in the phase diagrams of Figs.~\ref{fig:phase_diagram},
\ref{fig:phase_diagram_40} and \ref{fig:phase_diagram_100}
that the slopes of the three transition temperatures (i.e.
$dT_c/d(\mssq)$ at $\mssq=0$) are different, with that for the
$\De_3\rightarrow 0$ critical temperature the shallowest
and that for the $\De_2\rightarrow 0$ critical temperature
the steepest.  The ratios of these slopes can be
extracted from the Ginzburg-Landau free energy $\Omega$ of
(\ref{eq:GL}) and are given by 
\begin{equation}
1:(6\zeta-5):(6\zeta+4)
\end{equation}
where we have used $\beta_1=\beta_2$.
We can also read the slopes directly from our phase diagrams.
The $\De_0=25\MeV$ phase diagram of
Fig.~\ref{fig:phase_diagram} yields the ratios
\begin{equation}
1:10.1:19.1 
\end{equation}
which can be used to obtain 
two independent extractions of $\zeta$, one from
$1:(6\zeta-5)=1:10.1$ and the other from $1:(6\zeta+4)=1:19.1$. The
two are in perfect agreement, with both yielding 
$\zeta=2.52$ for $\De_0=25\MeV$.

\begin{figure}
\includegraphics[width=8cm]{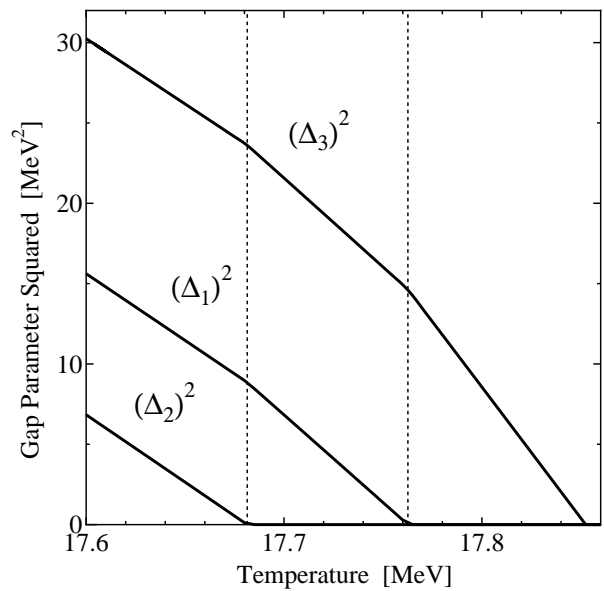}
\caption{Gap parameters squared versus $T$ at $\ms=30\MeV$, namely
$\mssq=1.8\MeV$.  
This figure should be compared to
the ``schematic illustration'' given in Fig.~2 of 
Ref.~\cite{Iida:2003cc}.  
We see the three phase transitions
separating the CFL, dSC, 2SC and unpaired phases.
}
\label{fig:gapsq}
\end{figure}

A third extraction can be obtained from Fig.~\ref{fig:gapsq} 
upon realizing that, according to (\ref{eq:GL}), in
the CFL phase where all three gap parameters are nonzero
$\zeta$ is given by the ratio
\begin{equation}
 \frac{\Delta_3^2-\Delta_2^2}{\Delta_1^2-\Delta_2^2}\ .
\label{eq:zetaratio}
\end{equation}
This ratio can be extracted at $\mssq=1.8\MeV$ from 
Fig.~\ref{fig:gapsq}. We have done this extraction
at a number values of $\mssq$ and fitted the results as
\begin{equation}
\frac{\Delta_3^2-\Delta_2^2}{\Delta_1^2-\Delta_2^2}
=2.52 +36.2\biggl(\frac{\ms}{\mu}\biggr)^2
    \!\!\!+1.02\!\times\!10^3\biggl(\frac{\ms}{\mu}\biggr)^4
\label{eq:thirdextraction}
\end{equation}
which leads us to conclude for the third time that 
$\zeta=2.52$ for $\De_0=25\MeV$.  In (\ref{eq:thirdextraction})
we have extracted the $M_s^2/\mu^2$ and $M_s^4/\mu^4$
corrections to the ratio (\ref{eq:zetaratio}), in addition
to extracting $\zeta$.  The coefficients we have obtained
confirm that these higher order terms are negligible
as long as $M_s^2/\mu^2\ll T_c/\mu$ or, equivalently given (\ref{eq:CFLTc}),
$\mssq\ll \De_0$.  This indicates yet again that the Ginzburg-Landau
free energy (\ref{eq:GL}) provides a good approximation 
to our results in the regime where it should be valid.

Upon comparing
our Fig.~\ref{fig:gapsq} with the schematic illustration
given in Fig.~2 of Ref.~\cite{Iida:2003cc}, there can already be
little doubt that the Ginzburg-Landau potential (\ref{eq:GL})
correctly describes the results that we have obtained by
solving the full gap and neutrality equations numerically, 
for $T\sim T_c$ and for $M_s^2/\mu^2\ll T_c/\mu$.  
The agreement between
our three extractions of $\zeta$ makes this point quantitatively.

We can also use Fig.~\ref{fig:gapsq} to check that our numerical
results are consistent with $\beta_1=\beta_2$.  The
Ginzburg-Landau potential (\ref{eq:GL}) can first be used to show
that in the CFL (and dSC) phases in Fig.~\ref{fig:gapsq}
the slopes of the lines for the three (two) nonzero gap
parameters are the same.  Our results clearly satisfy this.
Next, the Ginzburg-Landau potential (\ref{eq:GL}) can be used
to show that the ratio of the
slopes of the gap parameters in the CFL phase in Fig.~\ref{fig:gapsq}
to those in the dSC phase is  $(2\beta_1+\beta_2)/(3\beta_1+\beta_2)$,
and the ratio of those in the dSC phase to that in the 2SC phase
is $(\beta_1+\beta_2)/(2\beta_1+\beta_2)$.
These ratios cannot be extracted very accurately
from Fig.~\ref{fig:gapsq}, given the narrow windows
within which the dSC and 2SC phases are found. However, they
are consistent with $3/4$ and $2/3$, 
corresponding to $\beta_1=\beta_2$.

\begin{figure}[t]
\includegraphics[width=4cm]{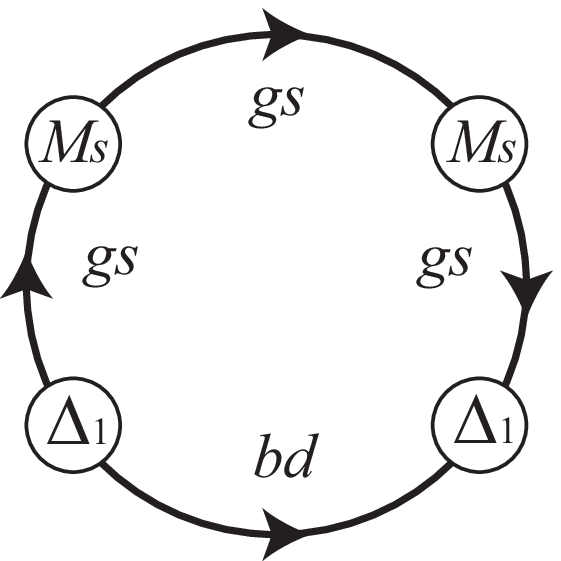}
\includegraphics[width=3.7cm]{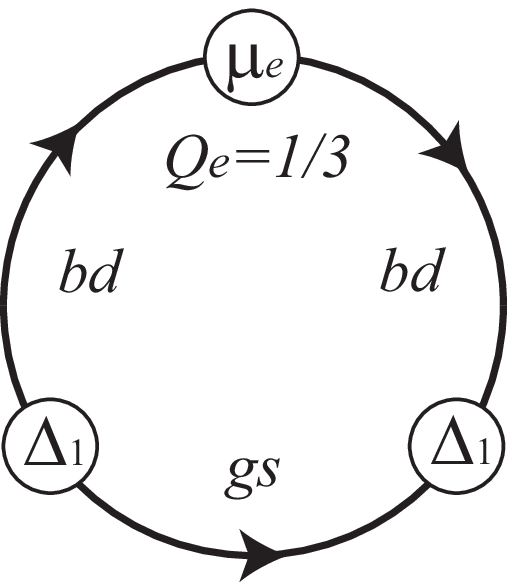}
\caption{Diagrams contributing to the $\epsilon$ and $\eta$ terms
in the Ginzburg Landau potential (\ref{eq:GL}).}
\label{fig:diagrams_ms_mue}
\end{figure}

In Ref.~\cite{Iida:2003cc}, the ratio $\zeta$ is calculated
using weak coupling methods, valid at asymptotic densities.
The weak coupling result is $\zeta=2$.  From our numerical
results, we have found $\zeta=2.52$ at $\De_0=25\MeV$.
We also find $\zeta=2.69$ at $\De_0=40\MeV$ and $\zeta=3.87$ 
at $\De_0=100\MeV$, extracting $\zeta$ from the ratio 
(\ref{eq:zetaratio}) as in (\ref{eq:thirdextraction}).  
At weak
coupling, $\zeta=2$ and
the $\De_3^2$ and $\De_2^2$ lines are equidistant
from the $\De_1^2$ line in the CFL phase region of Fig.~\ref{fig:gapsq}.
As the coupling gets stronger, the 
$\De_1^2$ and $\De_2^2$ lines move downward/leftward,
further away from the $\De_3^2$  line, and the ratio
$\zeta$ increases.

Now that we are convinced that the $\epsilon$ and $\eta$ terms
fully describe the $M_s$-dependent physics in our model
at small $M_s$, we calculate $\epsilon$ and $\eta$, from diagrams
like those in Fig.~\ref{fig:diagrams_ms_mue}. After some
calculation, the result is
\begin{widetext}
\begin{align}
\epsilon=&\frac{M_s^2}{4\pi^2\mu}\int_0^\Lambda dp \, p^2 \Biggl\{
\frac{1}{(p-\mu)^2}\tanh\left(\frac{p-\mu}{2T_c}\right)
-\frac{1}{(p+\mu)^2}\tanh\left(\frac{p+\mu}{2T_c}\right)\nonumber\\
&\qquad\qquad-\frac{\mu}{2T_c \,p (p-\mu)}\left[\cosh\left(\frac{p-\mu}{T_c}
\right)\right]^{-2}
-\frac{\mu}{2T_c\, p (p+\mu)}\left[\cosh\left(\frac{p+\mu}{T_c}
\right)\right]^{-2}
\Biggr\}
\label{eq:zeta_analytic1}
\\
\eta =&\frac{\mu_e}{2\pi^2}\int_0^\Lambda dp \, p^2 \Biggl\{
\frac{1}{(p-\mu)^2}\tanh\left(\frac{p-\mu}{2T_c}\right)
-\frac{1}{(p+\mu)^2}\tanh\left(\frac{p+\mu}{2T_c}\right)\nonumber\\
&\qquad\qquad-\frac{1}{2T_c(p-\mu)}\left[\cosh\left(\frac{p-\mu}{T_c}
\right)\right]^{-2}
+\frac{1}{2T_c (p+\mu)}\left[\cosh\left(\frac{p+\mu}{T_c}
\right)\right]^{-2}
\Biggr\}\ .
\label{eq:zeta_analytic2}
\end{align}
\end{widetext}
In the weak-coupling limit, $T_c\ll\mu$ and the integrals
are dominated by $p$ within $T_c$ of $\mu$.  In this limit,
it is easy to check that $\zeta=\epsilon/\eta=M_s^2/2\mu\mu_e$,
which yields $\zeta=2$ since $\mu_e=M_s^2/4\mu$.  At 
non-infinitesimal coupling, for example taking $\De_0=25\MeV$
and reading the corresponding $T_c$ from Fig.~\ref{fig:phase_diagram},
we can evaluate $\epsilon$ and $\eta$. We find $\zeta=2.55$
if we use $\mu_e=M_s^2/4\mu$, and if instead we
obtain $\mu_e$ from our numerical results,
we find $\zeta=2.52$.  Taking $T_c$ and $\mu_e$ from our
numerical results and evaluating $\zeta$ using 
(\ref{eq:zeta_analytic1},\ref{eq:zeta_analytic2}) 
we find $\zeta=2.52$, $2.65$, $3.84$
for $\De_0=25$, $40$, $100\MeV$ respectively.  From our numerical
results, we have found $\zeta=2.52$, $2.69$, $3.87$ 
at these values of $\De_0$, extracting $\zeta$ from the ratio 
(\ref{eq:zetaratio}) as in (\ref{eq:thirdextraction}).
The agreement between these determinations is a confirmation
of the accuracy of our numerical methods.

It is nice to see how quantitatively well the Ginzburg-Landau approximation
describes the physics near $T_c$ in our model, as we have demonstrated.
Furthermore, the value of one of the two ratios of coefficients that we have
investigated, $\beta_1/\beta_2=1$, is the same in our model
and in QCD at asymptotic densities.  And, the value
of the other ratio $\zeta=\eps/\eta$ is comparable in our model to 
its value in QCD at asymptotic densities for 
$\De_0=25$ and $40\MeV$, becoming significantly larger only for
quite strong coupling, as at $\De_0=100\MeV$.

\section{Implications and Open Questions}
\label{sec:conclude}

The phase diagrams shown in Figs.~\ref{fig:phase_diagram},
\ref{fig:phase_diagram_40} and \ref{fig:phase_diagram_100} constitute
the central results of this paper.  They could be used in
one of two different ways.  If future theoretical
advances constrain the $\mu$-dependent
values of $\De_0$ and $M_s$ more tightly than at present,
these phase diagrams (and those that interpolate between
them for other values of $\De_0$) could be used to construct
the phase diagram of nature.  Or, if future astrophysical
observations teach us that the phase diagram of nature must
have certain features, for example must or must not include a certain
phase, then the phase diagrams we have constructed 
could be used to draw inferences about the magnitudes of
$\De_0$ and $M_s$.

In thinking about the future phenomenological use of the phase diagrams
that we have found, their complexity raises concerns.
However, in most astrophysical contexts compact stars
have temperatures much less than 1 MeV.  At such low temperatures,
which can be thought of as $T=0$ in a QCD context, 
the phase diagrams are more manageable. We have the CFL
phase at asymptotic densities, with the gCFL phase
taking over at lower densities, when $\mu<M_s^2/2\De$.  
If $\De_0$ lies at the large end of its 
estimated range $10\MeV < \De_0 < 100 \MeV$,
it seems likely that hadronic matter will take over
from gCFL (or even from CFL) meaning that the complexities that
we have found in our phase diagrams 
at larger $\mssq$ will likely be superseded
by the transition to the hadronic phase.  

If $\De_0$ lies at the lower end of its allowed range,
then in Fig.~\ref{fig:phase_diagram} we find a straightforward
transition to ``unpaired'' quark matter as
the density is decreased further and $\mssq$ increases beyond
the gCFL window.  If we extend our pairing ansatz, we will
certainly find some pairing in this regime at sufficiently
low temperature. For example,
perhaps weak pairing between quarks with the same flavor
plays a role~\cite{Schmitt:2002sc,Alford:2002rz}, or perhaps
it is the crystalline color superconducting 
phase~\cite{Alford:2000ze,Casalbuoni:2004wm}
that takes over from gapless CFL at lower densities.
Recent developments~\cite{Casalbuoni:2004wm} make
the crystalline color superconducting phase look like
the most viable contender~\cite{Alford:2004hz}.

The one astrophysical context in which the full complexity
of the phase diagrams that we have analyzed must be faced
head-on is the physics of 
the proto-neutron star formed
during a supernova, and in particular of the propagation
of neutrinos therein.  Phenomena encoded in the phase
diagrams that we have analyzed could ultimately result
in observable consequences
in the time-of-arrival distribution of the neutrinos
detected from a future 
supernova~\cite{Carter:2000xf,Reddy:2002xc,Kundu:2004mz}.
The phenomenological implications of the 
complexity of the phase diagrams (with many phase
transition lines, the doubly critical point, the tricritical
point and the insulator to metal transition as the CFL phase
is heated) will have to be thought through
in this context.  The analytic treatment of the insulator
to metal crossover and the Ginzburg-Landau analysis of
the physics near $T_c$ for small $M_s^2/\mu T_c$ that
we have presented in Sections~\ref{sec:crossover} and \ref{sec:GL} 
could prove valuable
in this context.

Our analysis leaves 
many open avenues of investigation that must still be followed
to their conclusions.  The effects if 
gauge field fluctuations must be 
investigated, as must those of $K^0$-condensation in the
gapless CFL phase.  The effects of the 't~Hooft interaction
should be included, and the quark masses should be treated
as dynamical condensates to be solved for, rather than as parameters.
The pairing ansatz should be generalized, for example
to allow for a comparison between the free energies of
the gapless CFL and crystalline phases in three-flavor QCD.

Finally, the stability of the gapless CFL phase needs to
be investigated further, along the lines 
of Ref.~\cite{Huang:2004bg}. These authors 
find that the g2SC phase is unstable, as is the 2SC phase
near the 2SC$\rightarrow$g2SC transition.  This instability
could reflect the known instability of the g2SC phase
with respect to a mixed phase of charged components~\cite{Bedaque:2003hi},
it could reflect an instability 
with respect to the crystalline color superconducting phase,
or it could reflect the existence of some inhomogeneous
phase yet to be discovered.  We know that the gCFL phase
is stable with respect to mixed phases,
except perhaps at the largest values of $\mssq$ where it
is found in our phase diagrams~\cite{Alford:2004hz,Alford:2004nf}. 
It seems likely that at these large values of $\mssq$ 
the gCFL phase may anyway be superseded by the crystalline 
phase~\cite{Casalbuoni:2004wm,Alford:2004hz}. 
This could be clarified either by a direct free energy
comparison of these two phases, or by a stability analysis
of the gCFL phase as in Ref.~\cite{Huang:2004bg}.

\begin{acknowledgments}
We acknowledge helpful conversations with M.~Alford, J.~Bowers, M.~Forbes,
T.~Hatsuda, M.~Huang, J.~Kundu, T.~Schaefer, I.~Shovkovy and M.~Tachibana.
C.~K.\ and K.~R.\ thank the Institute for Nuclear Theory in Seattle
for its hospitality. K.~F.\ is grateful to D.~Rischke for
hospitality at the University of Frankfurt
during the completion of this work.
K.~F.\ was supported by Japan Society for the Promotion of Science for
Young Scientists. This research was supported in part by the U.S.\
Department of Energy (D.O.E.) under cooperative research agreement
\#DF-FC02-94ER40818.
\end{acknowledgments}

\end{document}